\documentclass[twocolumn,letterpaper,prb,superscriptaddress,showpacs,amsmath]{revtex4}

\usepackage{graphicx}

\newcommand{\im}{\mathrm i}
\newcommand{\vc}[1]{\boldsymbol{#1}}

\newcommand{\tpk}{t^{\phantom{.}}_{\perp{\vc k}_\parallel}}
\newcommand{\tppk}{t'_{\perp{\vc k}_\parallel}}

\begin{document}

\title{Microscopic gauge-invariant theory of the c-axis infrared response of
bilayer cuprate superconductors and the origin of the superconductivity induced absorption bands}

\author{Ji\v{r}\'{\i} Chaloupka}
\email[Electronic address:]{chaloupka@physics.muni.cz}
\affiliation{Department of Condensed Matter Physics, Faculty of Science,
Masaryk University, Kotl\'a\v{r}sk\'a 2, 61137 Brno, Czech Republic}

\author{Christian Bernhard}
\affiliation{Department of Physics and Fribourg Center for Nanomaterials, 
Chemin du Mus\`{e}e 3, CH-1700 Fribourg, Switzerland}

\author{Dominik Munzar}
\affiliation{Department of Condensed Matter Physics, Faculty of Science,
Masaryk University, Kotl\'a\v{r}sk\'a 2, 61137 Brno, Czech Republic}

\begin{abstract} 
We report on results of our theoretical study of the $c$-axis
infrared conductivity of bilayer high-$T_c$ cuprate superconductors using a
microscopic model involving the bilayer-split (bonding and antibonding) bands.
An emphasis is on the gauge-invariance of the theory, which turns out to be
essential for the physical understanding of the electrodynamics of these
compounds.  The description of the optical response involves local
(intra-bilayer and inter-bilayer) current densities and local conductivities.
The local conductivities are obtained using a microscopic theory, where the
quasiparticles of the two bands are coupled to spin fluctuations. The coupling
leads to superconductivity and is described at the level of generalized
Eliashberg theory.  Also addressed is the simpler case of quasiparticles
coupled by a separable and nonretarded interaction.  The gauge invariance of
the theory is achieved by including a suitable class of vertex corrections.
The resulting response of the model is studied in detail and an interpretation
of two superconductivity-induced peaks in the experimental data of the real
part of the $c$-axis conductivity is proposed. The peak around
$400\:\mathrm{cm}^{-1}$ is attributed to a collective mode of the
intra-bilayer regions, that is an analogue of the Bogolyubov-Anderson mode
playing a crucial role in the theory of the longitudinal response of
superconductors.  For small values of the bilayer splitting, its nature is
similar to that of the transverse plasmon of the phenomenological Josephson
superlattice model.  The peak around $1000\:\mathrm{cm}^{-1}$ is interpreted
as a pair breaking-feature that is related to the electronic coupling through
the spacing layers separating the bilayers.
\end{abstract}

\date{\today}


\maketitle

\section{INTRODUCTION}

The $c$-axis infrared response of the high-$T_{c}$ cuprate superconductors
(HTCS) is strongly sensitive to doping.
\cite{Homes:1995:PhysicaC,Schuetzman:1995:PRB,Bernhard:1999:PRB}  For
underdoped HTCS, it reveals a surprisingly weak coupling between adjacent unit
cells\cite{Tamasaku:1992:PRL} and a pronounced pseudogap.
\cite{Homes:1993:PRL}  In optimally doped materials, the real part of the
normal state (NS) conductivity $\sigma_{c}$ is almost frequency- and
temperature-independent for a broad range of frequencies and temperatures.
\cite{Bernhard:1999:PRB}  In contrast, the response of overdoped HTCS exhibits
a metallic behavior.\cite{Bernhard:1999:PRB}  These findings, in particular
the pseudogap, and the qualitative nature of the changes across the phase
diagram, make the $c$-axis response one of the most interesting properties of
the HTCS (for a review see Ref.~\onlinecite{Basov:2005:RMP}).  In materials
with two copper-oxygen planes per unit cell (the so called bilayer compounds),
the $c$-axis response also reflects the electronic coupling within the pair of
closely-spaced planes, that is of high interest for the following reasons: 
(\textit{i})~Its renormalization with respect to the noninteracting case is an
important fingerprint of the electronic correlations of the ground state.
(\textit{ii})~For underdoped HTCS, the manifestations of the pseudogap in
$\sigma_{c}$ interfere with those of the coupling.  A prerequisite for an
understanding of the $c$-axis pseudogap is thus a disentanglement of the
former from the latter. 
(\textit{iii})~The coupling may contribute to the condensation energy (see
Refs.~\onlinecite{Chakravarty:2004:Nature}, \onlinecite{Munzar:2003:PRB} and
references therein). 

The character of the coupling has been debated since the early years of
the high-$T_{c}$ research.  According to the conventional band theory, the
hopping between the planes should lead to a splitting of the conduction band
into two branches: a bonding branch corresponding to states that are symmetric
with respect to the mirror plane in the middle of the bilayer unit, and an
antibonding branch corresponding to states that are
antisymmetric.\cite{Andersen:1995:JPCS}  For some regions of the Brillouin
zone (BZ), the bonding band is expected to be located below the Fermi level
and the antibonding band above, which should give rise to the interband
transitions.\cite{Mazin:1992:PRB}

The experimental normal state (NS) infrared spectra of the bilayer compounds,
however, do not contain any structure that could be easily attributed to the
transitions.  Furthermore, the 20th-century photoemission experiments did not
reveal the splitting of the conduction band.  These findings could be
interpreted in terms of strong electronic correlations localizing charged
quasiparticles in individual planes, even in the case of the bilayer unit, and
inhibiting the band splitting. The simple band-structure based picture of the
NS thus seemed to have failed.  The experimental superconducting (SC) state
infrared spectra of underdoped bilayer compounds exhibit features that are
almost certainly related to the bilayer coupling: a broad absorption peak in
the spectra of $\mathrm{Re}\,\sigma_{c}$ in the frequency region between
$350\:\mathrm{cm}^{-1}$ and $550\:\mathrm{cm}^{-1}$ (labeled as $P_{1}$
in the following) and related anomalies of some infrared active
phonons.\cite{Homes:1995:PhysicaC,Schuetzman:1995:PRB,Munzar:1999:SSC} These
features, however, also appear to be consistent with the absence of the
conduction-band splitting and the localization of charged quasiparticles: It
was shown that they can be well understood and in some cases even fitted
\cite{Grueninger:2000:PRL,Munzar:1999:SSC} using the phenomenological model,
where the stack of the copper-oxygen planes is represented by a superlattice
of inter- and intra-bilayer Josephson junctions (the so called Josephson
superlattice model, JSM).\cite{Marel:1996:CzJP}  The mode $P_{1}$ has been
attributed to the transverse plasma mode of the model. A microscopic
justification of the model in terms of quasiparticle Green's functions has
been provided by Shah and Millis \cite{Shah:2001:PRB}. 

In the beginning of the 21st century, the situation changed.  In particular,
several groups have reported observations of two separate conduction bands in
photoemission spectra.
\cite{Feng:2001:PRL,Chuang:2001:PRL,Kordyuk:2002:PRL,
Borisenko:2002:PRB,Borisenko:2006:PRL}
The JSM is obviously not consistent with this observation.  In addition, it
became clear that the SC-state spectra of $\mathrm{Re}\,\sigma_{c}$ of
YBa$_{2}$Cu$_{3}$O$_{7-\delta}$ (Y-123) exhibit two distinct
superconductivity-induced modes: the mode $P_{1}$ discussed above and another
one around $1000\:\mathrm{cm}^{-1}$ (to be labeled as $P_{2}$).
\cite{Munzar:2003:PRB,Yu:2008:PRL}  It has been proposed, that the two are
related, but in the light of the results of the recent systematic
study by Yu {\it et al.} \cite{Yu:2008:PRL} this appears to be unlikely.  The
presence of $P_{2}$ cannot be accounted for in terms of the JSM.  These facts
thus call for a replacement of the simple phenomenological JSM  with a more
sophisticated theory involving the bilayer-split bands.  Here we present such
a theory and provide a fully microscopic interpretation of the
superconductivity-induced modes $P_{1}$ and $P_{2}$.

The basic ingredients of the theory are: 
(\textit{i})~The local current densities, conductivities, fields, and a
generalized multilayer formula.  The local current densities of the intra- and
inter-bilayer regions are expressed in terms of local conductivities and local
fields.   The fields differ from the average field because of charge
fluctuations between the planes.  Macroscopic considerations of these charging
effects lead to a formula for the total $c$-axis conductivity, that represents
an extension of the common multilayer formula \cite{Marel:1996:CzJP}.  
(\textit{ii})~The local conductivities are calculated using a microscopic
model and the linear response theory.  This is the main difference with
respect to the phenomenological JSM, where they are estimated or obtained by
fitting the data.  
(\textit{iii})~The microscopic description involves the two bilayer-split
bands.  The relevance of the bilayer splitting to the interpretation of the
$c$-axis response has been pointed out in Ref.~\onlinecite{Dordevic:2004:PRB}.
(\textit{iv})~The charged quasiparticles of the two bands are coupled to spin
fluctuations.  The coupling is treated at the level of generalized Eliashberg
theory, as in Ref.~\onlinecite{Chaloupka:2007:PRB}.  
(\textit{v})~The gauge invariance of the theory, required for a consistent,
\textit{i.e.}, charge conserving description of the charging effects, has been
achieved by including a class of vertex corrections (VC) ensuring that the
renormalized current vertices satisfy the appropriate Ward identities.  The
vertex corrections will be shown to lead to dramatic and qualitative changes
of the calculated response, similar to those occurring in case of the
longitudinal response of a homogeneous superconductor.

Calculated spectra of $\mathrm{Re}\,\sigma_{c}$ allow us to understand the
nature of the peaks $P_{1}$ and $P_{2}$.  The former will be shown to
correspond to a collective mode resembling the Bogolyubov-Anderson mode of
homogeneous superconductors and the latter to a pair breaking
(bonding-antibonding) peak.

The rest of the paper is organized as follows.  In Sec.~\ref{secTH} we present
the essential aspects of the theory, the values of the input parameters and
some computational details.  Section~\ref{secRD} contains results and
discussion.  In Sec.~\ref{secBCS} we focus on the relatively simple case of a
BCS-like interaction between the quasiparticles.  The analysis allows one to
understand the consequences of the bilayer splitting and the role of the
vertex corrections, but the resulting spectra of $\mathrm{Re}\,\sigma_{c}$ are
not sufficiently realistic.  The complex case of quasiparticles coupled to
spin fluctuations is addressed in Sec.~\ref{secEliashberg}.  It will be shown
that the calculated SC-state spectra display two distinct modes, similar to
the experimental ones.  Section~\ref{secCompofTandE} presents a comprehensive
discussion of the relation between theory and experiment including the
interpretation of the superconductivity-induced modes.  The summary and
conclusions are given in Sec.~\ref{secSM}.  The readers interested only in the
main findings of the paper may consider skipping Sec.~\ref{secTH}, and some
technical parts of Sec.~\ref{secBCS} and Sec.~\ref{secEliashberg}. 

\section{THEORY}
\label{secTH}

In this section we elaborate on the basic ingredients of our theory mentioned
in the introduction. First we briefly describe a phenomenological approach to
the $c$-axis electrodynamics of the bilayer systems.  In the subsequent
paragraphs, we build up a corresponding microscopic description. 

\subsection{Multilayer model}
\label{secMulti}

\begin{figure}[tbp]
\includegraphics{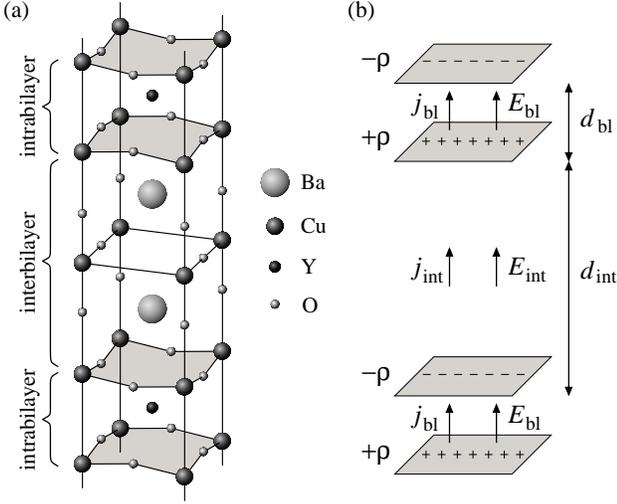}
\caption{(a) Crystal structure of Y-123. (b) Multilayer model, where
intrabilayer and interbilayer current densities $j_\mathrm{bl}$ and
$j_\mathrm{int}$ lead to a charge redistribution between the CuO$_2$ planes,
which modifies the local fields $E_\mathrm{bl}$ and $E_\mathrm{int}$.}
\label{fig:struct}
\end{figure}

The multilayer model proposed by van der Marel and Tsvetkov
\cite{Marel:1996:CzJP} provides a phenomenological description of the
\mbox{$c$-axis} electrodynamics of bilayer cuprates.  These compounds are
considered as consisting of homogeneously charged copper-oxygen planes
separated by intrabilayer (bl) and interbilayer (int) spacing regions (see
Fig.~\ref{fig:struct}). The dielectric function of the intrabilayer region 
\begin{equation}
\varepsilon_\mathrm{bl}(\omega) = \varepsilon_\infty + 
 \frac{i\sigma_\mathrm{bl}(\omega)}{\varepsilon_0\omega} \;,
\end{equation}
contains the interband dielectric constant $\varepsilon_\infty$ and the local
conductivity $\sigma_\mathrm{bl}$ defined by
$j_\mathrm{bl}=\sigma_\mathrm{bl}E_\mathrm{bl}$, where $j_\mathrm{bl}$ is the
local current density and $E_\mathrm{bl}$ the local field.  The interbilayer
region is described in a similar way using the local conductivity
$\sigma_\mathrm{int}$. To obtain the macroscopic (total) $c$-axis dielectric
function $\varepsilon(\omega)$, modifications of the local fields due to the
charging of the planes have to be considered. The result is
\begin{equation}\label{eq:vdMT}
\frac{d}{\varepsilon(\omega)}= 
 \frac{d_\mathrm{bl}}{\varepsilon_\mathrm{bl}(\omega)}+
 \frac{d_\mathrm{int}}{\varepsilon_\mathrm{int}(\omega)} \;.
\end{equation}

An extended version of the model, that we use in this paper, includes 
the dependence of the local current densities on both local fields:
\begin{equation}\label{eq:siglocphen}
j_L = \sum_{L'} \sigma_{LL'} E_{L'}\;, \quad 
L,L'\in \{\mathrm{bl},\mathrm{int}\} \;.
\end{equation}
The total $c$-axis conductivity $\sigma_c(\omega)$ 
is given as the ratio of the average current density 
$\langle j\rangle=(d_\mathrm{bl}j_\mathrm{bl}+d_\mathrm{int}j_\mathrm{int})/d$
to the average electric field 
$\langle E\rangle=(d_\mathrm{bl}E_\mathrm{bl}+d_\mathrm{int}E_\mathrm{int})/d$.
By employing the continuity relation between the charge and current densities
$j_\mathrm{int}-j_\mathrm{bl}=\partial\rho/\partial t$
and the effect of the charged planes on the local fields,
$E_\mathrm{bl}-E_\mathrm{int}=\rho/\varepsilon_0\varepsilon_\infty$
(see Fig.~\ref{fig:struct}), we arrive at 
\begin{equation}\label{eq:vdMT2}
\sigma_c(\omega)=
\frac{d_\mathrm{bl}\sigma_\mathrm{bl/bl}+d_\mathrm{int}\sigma_\mathrm{int/bl}}
      {d_\mathrm{bl}+d_\mathrm{int}\alpha}
+\frac{d_\mathrm{bl}\sigma_\mathrm{bl/int}+d_\mathrm{int}\sigma_\mathrm{int/int}}
      {d_\mathrm{bl}\alpha^{-1}+d_\mathrm{int}} \;,
\end{equation}
where
\begin{equation}
\alpha=\frac{E_\mathrm{int}}{E_\mathrm{bl}}
=\frac{\sigma_\infty+\sigma_\mathrm{bl/bl}-\sigma_\mathrm{int/bl}}
      {\sigma_\infty+\sigma_\mathrm{int/int}-\sigma_\mathrm{bl/int}}
\end{equation}
and $\sigma_\infty=-i\omega\varepsilon_0\varepsilon_\infty$.
The total dielectric function is given by
$\varepsilon(\omega)=\varepsilon_\infty+i\sigma_c(\omega)/\varepsilon_0\omega$.

In the following we describe the calculations of the local conductivities
$\sigma_{LL'}$ based on a microscopic model. The subsequent incorporation of
the interplane Coulomb interaction will then provide a microscopic
justification for the phenomenological treatment of the plane-charging effects
used in the model of van der Marel and Tsvetkov.

\subsection{Electronic structure -- tight-binding bands,
their renormalization and superconductivity}
\label{secBand}

One of the main components of our microscopic calculations are the two
bilayer split bands. We therefore begin with the tight-binding description of
these bands. The usual form of the in-plane dispersion
\begin{equation}\label{eq:band}
\epsilon_{{\vc k}_\parallel}=
 -2t(\cos k_xa+\cos k_ya)-4t'\cos k_xa\cos k_ya 
\end{equation}
will be considered, with the nearest neighbor and second nearest neighbor
hopping matrix elements $t$ and $t'$.  The intrabilayer hopping is governed by
the hopping matrix element $\tpk$ that is assumed to depend on $k_x$ and $k_y$
as
\begin{equation}\label{eq:tperp}
\tpk=\frac{t_{\perp\mathrm{max}} }4(\cos k_xa-\cos k_ya)^2 \;.
\end{equation}
This approximate form is suggested by the results of LDA calculations
\cite{Andersen:1995:JPCS} and is roughly consistent with experimental data on
\mbox{Bi$_2$Sr$_2$CaCu$_2$O$_{8+\delta}$} (Bi-2212) \cite{Feng:2001:PRL}.  Let
us note that the essential results of our calculations do not depend on the
form of $t_{\perp{\vc k}_\parallel}$, what matters is the magnitude.  In
addition to the intrabilayer hopping, we consider a weak coupling through the
interbilayer region with the matrix element $\tppk$ of the same $\vc
k$-dependence as $\tpk$.  The interlayer hopping splits the band
\eqref{eq:band} into two bands -- bonding (B) and antibonding (A) -- with the
dispersions
\begin{equation}
\epsilon_{B/A {\vc k}} = \epsilon_{{\vc k}_\parallel} \mp
\sqrt{t^2_{\perp{\vc k}_\parallel}+
      t'^2_{\perp{\vc k}_\parallel}+2\tpk\tppk\cos k_z d}\;.
\end{equation}

To account for the renormalization of charged quasiparticles and the
superconducting pairing we adopt the spin-fermion model,
where the quasiparticles are coupled to spin fluctuations.
In the case of a single band, the model selfenergy ($2$ by $2$ matrix)
is given by
\begin{equation}\label{eq:SFselfE}
\Sigma({\vc k},\im E)=\frac{g^2}{\beta N}\sum_{{\vc k'}, \im E'}
\chi_\mathrm{SF}({\vc k}-{\vc k'},\im E-\im E')\,
\mathcal{G}({\vc k'},\im E') \;,
\end{equation}
which can be schematically written as the convolution
$\Sigma=g^2 \chi_\mathrm{SF} \star \mathcal G$. Here $g$ is the coupling
constant, $\chi_\mathrm{SF}$ is the Matsubara counterpart of the spin
susceptibility and $\mathcal{G}$ the Nambu propagator, 
$\mathcal{G}(\vc k,iE)=
 [iE\tau_0-(\epsilon_{\vc k}-\mu)\tau_3-\Sigma(\vc k,iE)]^{-1}$.
\begin{figure}[tbp]
\includegraphics{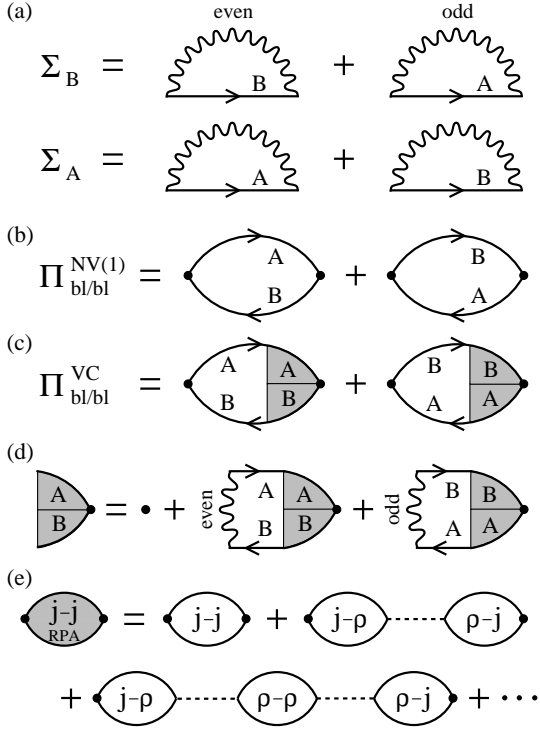}
\caption{(a) Diagrammatic representation of the selfenergies of the bonding 
(B) and antibonding (A) bands. The
propagators of the electronic quasiparticles and spin fluctuations are
represented by the straight and the wiggly lines respectively.
(b) Simple bubble approximation to the current-current correlator
\eqref{eq:jjcorr}. Only the part given by Eq.~\eqref{eq:jjsimple} is shown.
The black dots are the current vertices corresponding to $j_\mathrm{bl}^p$.
(c) Current-current correlator with a renormalized current
vertex \eqref{eq:jjvertex}. The diagrams corresponding to the case of $\tppk=0$ 
with no intraband contributions are shown.
(d) Diagrammatic representation of the Bethe-Salpeter equation \eqref{eq:BSE}.
(e) Diagrammatic representation of the equation determining the current-current
correlator including plane-charging effects. The dashed lines correspond to the 
interplane Coulomb interaction.
}
\label{fig:diagrams}
\end{figure}
The~generalization to the two band case is straightforward and the 
selfenergies can be expressed as\cite{Liechtenstein:1996:PRB,Eschrig:2002:PRL}
\begin{equation}\label{eq:SFselfEAB}
\Sigma_{B/A}=g^2 \chi_\mathrm{SF}^\mathrm{odd}\star \mathcal{G}_{A/B}
        + g^2 \chi_\mathrm{SF}^\mathrm{even}\star \mathcal{G}_{B/A} \;,
\end{equation}
where we distinguish between the spin-susceptibility channels of even
($\chi_\mathrm{SF}^\mathrm{even}$) and odd ($\chi_\mathrm{SF}^\mathrm{odd}$)
symmetry with respect to the mirror plane in the
center of the bilayer unit. The diagrammatic representation of $\Sigma_{B/A}$
is shown in Fig.~\ref{fig:diagrams}(a).  We have used the same form of
$\chi_\mathrm{SF}$ containing the resonance mode and a broad continuum as in
Ref.~\onlinecite{Chaloupka:2007:PRB} (details will be given in
Sec.~\ref{secParam}).  

Since the results of the selfconsistent calculations based on the spin-fermion
model are difficult to interpret, we first resort to the BCS level.  The
results obtained this way are easier to understand because of the absence of
retardation and better possibilities of analytical manipulations of the
formulas. The even/odd interaction channels are assumed to be equivalent
which leads to the same superconducting gap 
$\Delta_{\vc k}=\frac12\Delta_\mathrm{max}(\cos k_xa-\cos k_ya)$ 
in both bands determined by
\begin{equation}\label{eq:BCSselfE}
\Delta_{\vc k}=-\sum_{\vc k',n\in\{A,B\}} V_{\vc k \vc k'}
\frac{\Delta_{\vc k'}}{2E_{\vc k'n}}\tanh\frac{\beta E_{\vc k' n}}2 \;,
\end{equation}
where $V_{\vc k\vc k'}=-\lambda w_{\vc k}w_{\vc k'}$ with
$w_{\vc k}=(\cos k_xa-\cos k_ya)/2$ is the BCS interaction 
of $d$-wave symmetry and $E_{\vc k A/B}$ is the usual BCS quasiparticle 
energy $E_{\vc k A/B}=\sqrt{(\epsilon_{\vc k A/B}-\mu)^2+\Delta_{\vc k}^2}$.
For details see Appendix.

\subsection{Response to electromagnetic field}
\label{secResp}

Here we calculate the response of the model to the $c$-axis
polarized electromagnetic field represented by the external vector potential
$\vc A_\mathrm{ext}=(0,0,A_\mathrm{ext})\,e^{i \vc q\cdot\vc R-i\omega t}$.
The coupling of the tight-binding model to the electromagnetic field can be
obtained by multiplying each hopping term by the corresponding 
Peierls phase factor according to the prescription
\cite{Peierls:1933:ZP,Luttinger:1951:PR,Scalapino:1992:PRL}
$c^\dagger_{\vc R}c^{\phantom{\dagger}}_{\vc R'}\rightarrow
\exp\left[ -(ie/\hbar) \vc A_\mathrm{ext}\cdot (\vc R-\vc R') \right] 
c^\dagger_{\vc R}c^{\phantom{\dagger}}_{\vc R'}$.
To fit the scheme of Sec.~\ref{secMulti}, we formally distinguish between
the vector potentials
$A_\mathrm{bl}$ and $A_\mathrm{int}$, used for the hopping processes through
the intrabilayer and interbilayer regions respectively.  By expanding to the
second order in the vector potentials, we arrive at the coupling Hamiltonian
that can be used for extracting the $c$-axis paramagnetic and diamagnetic
current density operators.\cite{Scalapino:1992:PRL} 
The paramagnetic current density for $\vc q = 0$,
averaged over the corresponding region (bl/int), can be expressed as
\begin{equation}\label{eq:jpbl}
\begin{split}
\hat{\jmath}_\mathrm{bl/int}^{\,p} = -\frac{ie}{Na^2\hbar} 
\sum_{{\vc k}_\parallel k_z s} 
& \left[ \pm J^{(1)}_{\mathrm{bl/int},\vc k}
\left(c^\dagger_{A{\vc k}s}c^{\phantom{\dagger}}_{B{\vc k}s}-
      c^\dagger_{B{\vc k}s}c^{\phantom{\dagger}}_{A{\vc k}s}\right) \right.\\
& \left. + J^{(2)}_{\vc k}
\left(c^\dagger_{A{\vc k}s}c^{\phantom{\dagger}}_{A{\vc k}s}-
      c^\dagger_{B{\vc k}s}c^{\phantom{\dagger}}_{B{\vc k}s}\right) \right]
\end{split}
\end{equation}
with the matrix elements
\begin{equation}
J^{(1)}_{\mathrm{bl},{\vc k}}=
\frac{2\tpk(\tpk+\tppk\cos k_zd)}{\epsilon_{A\vc k}-\epsilon_{B\vc k}}
\end{equation}
and
\begin{equation}
J^{(2)}_{\vc k}=\frac{2i\tpk\tppk\sin k_zd}{\epsilon_{A\vc k}-\epsilon_{B\vc k}} \;.
\end{equation}
The matrix element $J^{(1)}_{\mathrm{int},\vc k}$ is obtained 
from $J^{(1)}_{\mathrm{bl},\vc k}$ simply by interchanging $\tpk$ and $\tppk$.
In the $t'_{\perp{\vc k}_\parallel}=0$ case, where
$J^{(1)}_{\mathrm{bl},\vc k}=\tpk$ and
$J^{(1)}_{\mathrm{int},\vc k}=J^{(2)}_{\vc k}=0$,
we arrive at the simplified expression 
\begin{equation}
\hat{\jmath}_\mathrm{bl}^{\,p} = -\frac{ie}{N_\parallel a^2\hbar} 
\sum_{{\vc k}_\parallel s} \tpk
\left(c^\dagger_{A{\vc k}_\parallel s}
      c^{\phantom{\dagger}}_{B{\vc k}_\parallel s}-
      c^\dagger_{B{\vc k}_\parallel s}
      c^{\phantom{\dagger}}_{A{\vc k}_\parallel s}\right) \;.
\end{equation}
The summation runs over ${\vc k}_\parallel$ from the 2D Brillouin zone
only and $N$ is reduced accordingly. The diamagnetic current density 
is given by
\begin{multline}\label{eq:jdbl}
\hat{\jmath}_\mathrm{bl/int}^{\;d} = 
-\frac{e^2d_\mathrm{bl/int}A_\mathrm{bl/int}}{Na^2\hbar^2} 
\sum_{{\vc k}_\parallel k_z s} 
 \left[ J^{(1)}_{\mathrm{bl/int},\vc k}
 \left(n_{B{\vc k}s}-n_{A{\vc k}s}\right) \right. \\ 
 \left. \mp J^{(2)}_{\vc k}
\left(c^\dagger_{A{\vc k}s} c^{\phantom{\dagger}}_{B{\vc k}s}-
      c^\dagger_{B{\vc k}s} c^{\phantom{\dagger}}_{A{\vc k}s}\right) \right]\;.
\end{multline}
In the $t'_{\perp{\vc k}_\parallel}=0$ case, Eq.~\eqref{eq:jdbl} simplifies to 
\begin{equation}
\hat{\jmath}_\mathrm{bl}^{\;d} = 
-\frac{e^2d_\mathrm{bl}A_\mathrm{bl}}{N_\parallel a^2\hbar^2} 
\sum_{{\vc k}_\parallel s} \tpk
\left(n_{B{\vc k}s}-n_{A{\vc k}s}\right) \;.
\end{equation}

The total $c$-axis conductivity is constructed along the lines of
Sec.~\ref{secMulti}. To this end, the current densities induced by the
electric fields $E_L=i(\omega+i\delta) A_L$
($L\in\{\mathrm{bl},\mathrm{int}\}$) have to be calculated and the local
conductivities determined from $j_L(\vc q,\omega)=\sum_{L'}\sigma_{LL'}(\vc
q,\omega)E_{L'}(\vc q,\omega)$.  At this point, the fields $E_{L'}$ are still
equal to the \textit{external} field $E_\mathrm{ext}$. However, it will be
shown in Sec.~\ref{secRPA}, that the local conductivities calculated as
outlined above, ignoring the charging effects play exactly the same role as in
Eq.~\eqref{eq:siglocphen}, \textit{i.e.}, they represent the response to the
\textit{local} fields.  Within the framework of the linear response theory,
the local conductivities are given by the Kubo formula
\begin{equation}\label{eq:Kubo}
\sigma_{LL'}(\vc q,\omega)=
\frac{(e^2/\hbar^2)K_L\delta_{LL'}+\Pi_{LL'}(\vc q,\omega)}{i(\omega+i\delta)}
\;.
\end{equation}
The first term in the numerator,
\begin{equation}\label{eq:Efkinen}
K_\mathrm{bl/int} = -\frac{d_\mathrm{bl/int}}{Na^2}
\sum_{{\vc k}_\parallel k_z s}
 J^{(1)}_\mathrm{bl/int,\vc k} \langle n_{B{\vc k}s}-n_{A{\vc k}s}\rangle \;,
\end{equation}
comes from the diamagnetic current densities and is related to the $c$-axis
kinetic energy\cite{Scalapino:1992:PRL}: In the $\tppk=0$ case, 
$K_\mathrm{bl}=(d_\mathrm{bl}/a^2)\langle T\rangle$,
where $\langle T\rangle$ is the intrabilayer kinetic energy per unit cell,
$T=-(1/N_\parallel)\sum_{\vc k_\parallel s} t_{\perp \vc k_\parallel}
(n_{B\vc k_\parallel s}-n_{A\vc k_\parallel s})=
-(1/N_\parallel)\sum_{\vc R_\parallel\vc R'_\parallel s} 
t_{\perp \vc R_\parallel\vc R'_\parallel}
(c^\dagger_{2\vc R_\parallel s}c^{\phantom{\dagger}}_{1\vc R'_\parallel s}
+c^\dagger_{1\vc R_\parallel s}c^{\phantom{\dagger}}_{2\vc R'_\parallel s})$.
The second term in Eq.~\eqref{eq:Kubo} is the retarded
correlation function of the para\-magnetic current densities
\begin{equation}\label{eq:jjcorr}
\Pi_{LL'}(\vc q,\omega)=
 i\frac{Na^2d_{L'}}{\hbar} 
 \int\limits_{-\infty}^{\infty}\mathrm{d}t\, e^{i\omega t}
 \langle[\hat{\jmath}^{\;p}_L(\vc q,t),
         \hat{\jmath}^{\;p}_{L'}(-\vc q,0)]\rangle
 \theta(t) \;.
\end{equation}

In the simplest approximation, the correlator is obtained by evaluating the
bubble diagrams where the two current vertices are joined by two electron
propagator lines. This is the approximation, where the vertex corrections are
neglected. Since the propagators refer to the two bands, there are four
possible combinations in total. Two of them correspond to interband
transitions and their contribution to the Matsubara counterpart of
\eqref{eq:jjcorr} at $\vc q=0$ equals
\begin{multline}\label{eq:jjsimple}
\Pi_{LL'}^{\mathrm{NV}(1)}(\vc q=0,i\hbar\nu)=
\mp\frac{e^2}{\hbar^2} \frac{d_{L'}}{Na^2\beta} \sum_{{\vc k},iE} 
 J_{L,\vc k}^{(1)} J_{L',\vc k}^{(1)} \\
\times \mathrm{Tr}\left[
   \mathcal{G}_A(\vc k,iE+i\hbar\nu)\mathcal{G}_B(\vc k,iE) \right. \\
  +\left.\mathcal{G}_B(\vc k,iE+i\hbar\nu)\mathcal{G}_A(\vc k,iE)\right]
\end{multline}
with the minus sign for $L=L'$ and plus sign for $L\neq L'$. The
corresponding diagrams are presented in Fig.~\ref{fig:diagrams}(b).
For $\tppk\neq 0$, all the conductivity components acquire,
in addition, an intraband contribution given by
\begin{multline}\label{eq:jjintra}
\Pi_{LL'}^{\mathrm{NV}(2)}(\vc q=0,i\hbar\nu)=
-\frac{e^2}{\hbar^2} \frac{d_{L'}}{Na^2\beta} \sum_{{\vc k},iE} 
 J^{(2)}_{\vc k}J^{(2)}_{\vc k} \\
\times \mathrm{Tr}\left[
   \mathcal{G}_A(\vc k,iE+i\hbar\nu)\mathcal{G}_A(\vc k,iE) \right. \\
  +\left.\mathcal{G}_B(\vc k,iE+i\hbar\nu)\mathcal{G}_B(\vc k,iE)\right] \;.
\end{multline}
This contribution has a similar frequency dependence as the in-plane
conductivity, the main difference coming from the $\vc k$-dependence of the
matrix element $J^{(2)}_{\vc k}$.
Typically, it is rather small compared to \eqref{eq:jjsimple}.

\subsection{Vertex corrections}
\label{secVertex}

The well-known deficiency of the simple bubble approximations such as the one 
leading to Eqs.~\eqref{eq:jjsimple} and \eqref{eq:jjintra}
is the lack of the gauge invariance which manifests itself, \textit{e.g.}, by
a violation of the normal-state restricted sum rule for the conductivity.
For the normal state the conductivity components should satisfy
the sum rule
$\int_{0+}^\infty \mathrm{Re}\,\sigma_{LL}(\omega) \,\mathrm{d}\omega 
=-(\pi e^2/2\hbar^2) K_L$.
While the discrepancy between the left-hand side and the right-hand side in
the corresponding case of the in-plane response is rather small (of the order
of $1\%$\cite{Chaloupka:2007:PRB}), here it is quite detrimental -- typically
$20-30\%$ -- as demonstrated in Sec.~\ref{secRD}.  Since there is an
intimate relation between the gauge invariance of the response functions and
the charge conservation, the large discrepancy indicates, that the continuity
equation between the current and charge densities is not even approximately
satisfied. As a consequence, the use of the formula \eqref{eq:vdMT2}, which
relies on the continuity equation, becomes questionable. In the following
paragraph we show explicitly, how the requirement of gauge invariance enters a
microscopic derivation of the formulas of Sec.~\ref{secMulti}.

To avoid the problems mentioned above, a gauge-invariant extension of the
approximation \eqref{eq:jjsimple}+\eqref{eq:jjintra} is necessary.  As found
by Nambu\cite{Nambu:1960:PR}, the gauge invariance of the response function is
guaranteed if we replace the bare current-density vertex with a properly
renormalized one. The required renormalization of this vertex (\textit{i.e.}, of the
interaction of the quasiparticles with photons) is determined by the form of
the quasiparticle selfenergy via the generalized Ward
identity.\cite{Schrieffer:1988:Book}

Here the situation is complicated by the presence of the two bands.  To be
able to express all the contributions in a systematical way, we first introduce
the bare vertex factors $(ie/Na^2\hbar)\gamma_{nm}^L(\vc k)$ (with
$m,n\in\{A,B\}$) inferred from \eqref{eq:jpbl}.  In the corresponding diagram,
the $m$-th band propagator line with momentum $\vc k$ enters the current
vertex of $j_L^p$ and the $n$-th band propagator line leaves it. The possible
combinations are:
$\gamma_{AB}^\mathrm{bl}=-\gamma_{BA}^\mathrm{bl}=
J^{(1)}_{\mathrm{bl},\vc k}$,
$\gamma_{BA}^\mathrm{int}=-\gamma_{AB}^\mathrm{int}=
J^{(1)}_{\mathrm{int},\vc k}$,
$\gamma_{BB}^\mathrm{bl}=-\gamma_{AA}^\mathrm{bl}=
 \gamma_{BB}^\mathrm{int}=-\gamma_{AA}^\mathrm{int}=
J^{(2)}_{\vc k}$.
The correlator $\Pi_{LL'}$ involving the renormalized current vertices
$\Gamma_{nm}^{L}(\vc k,iE,i\hbar\nu)$
\begin{multline}\label{eq:jjvertex}
\Pi_{LL'}^\mathrm{VC}(\vc q=0,i\hbar\nu)=
\frac{e^2}{\hbar^2} \frac{d_{L'}}{Na^2\beta}\sum_{\vc k, iE, mn\in\{A,B\}}
\mathrm{Tr}\left[\gamma_{mn}^{L'}(\vc k) \right.\\
            \left.\times\mathcal{G}_m(\vc k,iE)
            \Gamma_{nm}^{L}(\vc k,iE,i\hbar\nu)
            \mathcal{G}_n(\vc k,iE+i\hbar\nu)\right]
\end{multline}
contains two interband contributions with $mn=AB$ and $mn=BA$. The
corresponding diagrams are shown in Fig.~\ref{fig:diagrams}(c). For
$\tppk\!=0$, these are the only contributions. In the \mbox{$\tppk\!\neq 0$}
case, also the intraband terms with $mn=AA$, $mn=BB$ contribute.

The renormalized vertices $\Gamma_{nm}^{L}(\vc k,iE,i\hbar\nu)$ consistent
with the electronic selfenergies of the two bands obey the Bethe-Salpeter
equations of the form diagrammatically shown in Fig.~\ref{fig:diagrams}(d).
At this point, we have to distinguish between the spin-fluctuation mediated
interaction and the BCS interaction allowing for further analytical
simplifications.  Evaluating the diagrams in the former case we arrive at
\begin{equation}\label{eq:BSE}
\begin{split}
\Gamma_{AB}^L(\vc k,&iE,i\hbar\nu)=\gamma_{AB}^L(\vc k)\:\tau_0+ \\
+\frac{g^2}{\beta N}& \sum_{\vc k',iE'} 
 \chi_\mathrm{SF}^\mathrm{even}(\vc k-\vc k',iE-iE') \\
  &\times\mathcal{G}_B(\vc k',iE')
  \Gamma_{AB}^L(\vc k',iE',i\hbar\nu)
  \mathcal{G}_A(\vc k',iE'+i\hbar\nu) \\
+\frac{g^2}{\beta N}& \sum_{\vc k',iE'} 
 \chi_\mathrm{SF}^\mathrm{odd}(\vc k-\vc k',iE-iE') \\
  &\times\mathcal{G}_A(\vc k',iE')
  \Gamma_{BA}^L(\vc k',iE',i\hbar\nu)
  \mathcal{G}_B(\vc k',iE'+i\hbar\nu) \;
\end{split}
\end{equation}
and similar equations for the other renormalized vertices.
Intraband current vertices $\Gamma_{AA}^L$ and $\Gamma_{BB}^L$
turn out to be simply the bare ones because of the symmetry of 
$\gamma_{AA}^L(\vc k)$, $\gamma_{BB}^L(\vc k)$ (odd functions of $\vc k_z$)
and $q_z$-independence of $\chi_\mathrm{SF}(\vc q,\omega)$ assumed in 
the \mbox{$\tppk\!\neq 0$} case.
In the BCS case, the interaction is non-retarded and separable, which leads
to a simple $\vc k$-dependence and $iE$-independence of $\Gamma$:
$\Gamma_{AB}^L(\vc k,i\hbar\nu)=
 \gamma_{AB}^L(\vc k)+\lambda w_{\vc k} C^L(i\hbar\nu)$.
Here $\lambda$ is the BCS coupling constant and $w_{\vc k}$ is the $d$-wave
symmetry function introduced in Sec.~\ref{secBand}. The Bethe-Salpeter 
equations and the current-current correlators can then be treated 
to a large extent analytically\cite{Schrieffer:1988:Book}, as shown in
Appendix.
In addition, the intraband contributions are exactly zero in the optical limit
of $\vc q\rightarrow 0$.

\subsection{RPA approximation of plane-charging effects}
\label{secRPA}

In paragraph \ref{secMulti}, we presented the results of a phenomenological
approach to the effects due to the charging of the planes.  Here
we outline a rigorous microscopic derivation of Eq.~\eqref{eq:vdMT}, where
these effects are treated at the level of the random-phase approximation (RPA).
For the sake of simplicity, we restrict ourselves to the case of insulating
interbilayer regions.

The current density within a bilayer unit leads to a redistribution of charge
among the CuO$_2$ planes. The electrostatic interaction of the corresponding
charge densities is given by the interaction Hamiltonian
\begin{equation}
\hat{H}_\mathrm{Coulomb} =
\frac{Na^2 d_\mathrm{bl}}{2\varepsilon_\infty\varepsilon_0}\,
\hat{\rho}\,\hat{\rho} \;,
\end{equation}
where $\hat\rho$ is the excess planar charge density. The current-current
correlator $\Pi_\mathrm{bl/bl}$ modified by this interaction along the lines
of the RPA approximation, corresponding to the diagrammatic series shown in
Fig.~\ref{fig:diagrams}(e), reads
\begin{equation}
\Pi^\mathrm{RPA}_\mathrm{bl/bl} = \Pi^{j-j} -\Pi^{j-\rho}
\frac{1}{\varepsilon_\infty\varepsilon_0+\Pi^{\rho-\rho}} \, \Pi^{\rho-j} \;,
\end{equation}
where $\Pi^{j-j}\equiv \Pi_\mathrm{bl/bl}$, $\Pi^{j-\rho}$, $\Pi^{\rho-j}$,
and $\Pi^{\rho-\rho}$ are the correlation functions obtained without
considering the charging effects.
To proceed further towards Eq.~\eqref{eq:vdMT}, we have to express these
correlation functions using
the conductivity-related current-current correlator $\Pi^{j-j}$ only,
eliminating $\Pi^{j-\rho}$, $\Pi^{\rho-j}$,
and $\Pi^{\rho-\rho}$. This can be achieved
using the continuity equation for the charge and current densities. 
Let us note, that the gauge invariance of the local response functions 
is the necessary condition for the continuity equation to be valid. 
The result of the elimination can be written as
\begin{equation}
\sigma^\mathrm{RPA}_\mathrm{bl/bl} = 
\frac{\sigma_\mathrm{bl/bl}}
{1+\dfrac{i\sigma_\mathrm{bl/bl}}{\varepsilon_\infty\varepsilon_0\omega}} \;.
\end{equation}

The last step is the incorporation of the macroscopic averaging to obtain the
macroscopic $c$-axis dielectric function
\begin{equation}\label{eq:epsavg}
\varepsilon(\omega) = \varepsilon_\infty +
\frac{i}{\varepsilon_0\omega} \frac{\langle j\rangle}{\langle E\rangle}\;,
\end{equation} 
where the symbols $\langle j\rangle$ and $\langle E\rangle$ denote the
unit-cell averages of the current density and the electric field, respectively.  The
averaged current density is given by $\langle j\rangle=(d_\mathrm{bl}/d)
j_\mathrm{bl}$, since the interbilayer regions are supposed not to contribute.
The macroscopic field $\langle E\rangle$ consists of the
homogeneous external field and the averaged field of the induced charge
density $\langle E\rangle=
E_\mathrm{ext}+(d_\mathrm{bl}/d)\,\rho/\varepsilon_0\varepsilon_\infty$.
Using the relation 
$j_\mathrm{bl}=\sigma^\mathrm{RPA}_\mathrm{bl/bl}E_\mathrm{ext}$
and the continuity equation $i\omega\rho=j_\mathrm{bl}$, we obtain
\begin{equation}
\langle j \rangle = \frac{d_\mathrm{bl}}d
\sigma^\mathrm{RPA}_\mathrm{bl/bl} E_\mathrm{ext}\;,\quad
\langle E\rangle = E_\mathrm{ext} -
\frac{d_\mathrm{bl}}d \frac{i\sigma^\mathrm{RPA}_\mathrm{bl/bl}}
{\varepsilon_\infty\varepsilon_0\omega} E_\mathrm{ext} \;.
\end{equation}
Finally, by inserting these results in Eq.~\eqref{eq:epsavg}, we arrive at
Eq.~\eqref{eq:vdMT} with $\varepsilon_\mathrm{int}=\varepsilon_\infty$ and
$\varepsilon_\mathrm{bl}=\varepsilon_\infty
+i\sigma_\mathrm{bl/bl}/\varepsilon_0\omega$.
The local response function $\sigma_\mathrm{bl/bl}$ calculated in
Sec.~\ref{secResp} and \ref{secVertex} plays the role of $\sigma_\mathrm{bl}$.
In the more general case of Eq.~\eqref{eq:vdMT2}, the derivation is analogous
to the one presented here.  We stress, that the use of Eq.~\eqref{eq:vdMT} or
\eqref{eq:vdMT2} is now accompanied by the requirement of the gauge invariance
of the local conductivities.

\subsection{Input parameters and computational details}
\label{secParam}

The values of most of the input parameters are the same as in
Ref.~\onlinecite{Chaloupka:2007:PRB}. For the description of the bands we use
the in-plane dispersion with $t=350\:\mathrm{meV}$, $t'=-100\:\mathrm{meV}$
and the band filling $n=0.82$.  The values of the interplane hopping
parameters will be specified later at the corresponding places in the text, 
since various regimes of the optical response corresponding to various values
of these parameters are discussed.
In the multilayer formula, we use $d_\mathrm{bl}=3.4\:\mathrm{\AA}$, 
$d_\mathrm{int}=12.0\:\mathrm{\AA}$, \textit{i.e.}, the values 
corresponding to Bi-2212, and $\varepsilon_\infty=5$.

The model spin susceptibility has the same form as in
Refs.~\onlinecite{Casek:2005:PRB} and \onlinecite{Chaloupka:2007:PRB}
containing the $40\:\mathrm{meV}$ resonance mode and a continuum with
dimensionless spectral weights of $0.01b_M$ and $0.01b_C$ respectively.  In the
$\tppk=0$ case, we distinguish between the channels of odd and even symmetry
and include the resonant mode with $b_M=1$ in the odd channel only. The
continuum with $b_C=2$ is present in both channels.  For $\tppk\neq 0$, the
bonding and antibonding states are no more of the simple form
$|B\rangle,|A\rangle=(|1\rangle\pm|2\rangle)/\sqrt2$, where $|1\rangle$ and
$|2\rangle$ are state vectors residing on the first and the second plane of
the bilayer unit, respectively. The linear combination now contains $\vc
k$-dependent coefficients.  A proper construction of the the interaction
vertices would extensively complicate the theory. To avoid this complexity, 
we take $b_M=1/2$ and $b_C=2$ for both channels whenever $\tppk>0$.

The coupling constant $g=3\:\mathrm{eV}$ was chosen to yield $T_c$ around
$90\:\mathrm{K}$ and the amplitude of the superconducting gap $\Delta$ around
$30\:\mathrm{meV}$. Some of the calculations were performed on the simpler
BCS level, where we choose the value of the BCS coupling constant $\lambda$ 
leading to the same gap amplitude of $30\:\mathrm{meV}$.

The selfconsistent equations for the selfenergies \eqref{eq:SFselfEAB} and
Bethe-Salpeter equations \eqref{eq:BSE} were solved iteratively using a
Brillouin zone grid of typically $64\times 64\times 32$ points, and a cutoff
of $8\:\mathrm{eV}$ in Matsubara frequencies.  In the case of small
$t_{\perp\mathrm{max}}\lesssim 50\:\mathrm{meV}$, the vertex corrections
lead to a complete change of the response-function profiles and up to $10^3$
iterations of the Bethe-Salpeter equation are required to achieve the
convergence.  The convolutions were performed using the FFT algorithm with the
use of the symmetries of $\Sigma$ and $\Gamma$. Since the calculations are
very demanding in terms of computer time and memory, we have used 
$q_z$-independent spin susceptibility which brings the advantage of 
$k_z$-independent $\Sigma$ and $\Gamma-\gamma$. The calculated response functions
were continued to the real axis using the method of Pad\'{e}
approximants.\cite{Vidberg:1977:JLTP}

\section{RESULTS AND DISCUSSION}
\label{secRD}

\subsection{Quasiparticles paired by the BCS interaction}
\label{secBCS}

\begin{figure}[tbp]
\includegraphics{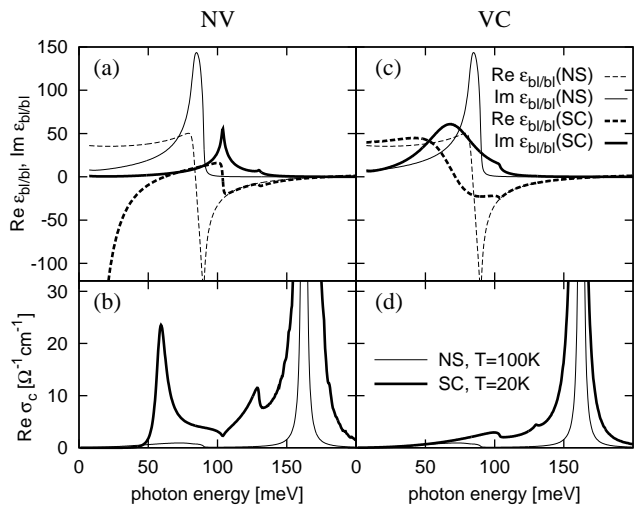}
\caption{(a) Local dielectric function $\varepsilon_\mathrm{bl/bl}$ 
in the simplest BCS case with $\tppk=0$, $t_{\perp\mathrm{max}}=45\:\mathrm{meV}$, 
and  $\Delta_\mathrm{max}=30\:\mathrm{meV}$,
vertex corrections are not included (NV). 
The thin (thick) lines correspond to the normal (superconducting) state,
$T=100\:\mathrm{K}$ 
($T=20\:\mathrm{K}$). 
The solid (dashed) lines represent the imaginary (real) part.  
(b) The real part of the corresponding total $c$-axis conductivity 
obtained using Eq.~\eqref{eq:vdMT}. 
The thin (thick) line corresponds to the normal (superconducting) state. 
(c,d) The same as in (a,b) but with the VC included (for the superconducting state only).} 
\label{fig:BCStp01}
\end{figure}

We begin with the simpler case of insulating spacing layers, 
\textit{i.e.}, $t'_{\perp}=0$. Figure \ref{fig:BCStp01}(a) shows 
the local dielectric function $\varepsilon_\mathrm{bl/bl}$ of the intrabilayer
region obtained using the bubble diagram of Fig.~\ref{fig:diagrams}(b),
\textit{i.e.},
with the vertex corrections neglected 
(this is abbreviated as NV).  
The thin (thick) lines
correspond to the normal (superconducting) state, the solid (dashed) lines
represent the imaginary (real) part.  
The NS response exhibits a sharp
absorption band near $80\:\mathrm{meV}$ due to the interband
(bonding-antibonding) transitions.  The SC-state response involves the
superconducting condensate, which manifests itself in the real part of
$\varepsilon_\mathrm{bl/bl}$ and a pair breaking peak at $110\:\mathrm{meV}$
corresponding to final states with one Bogolyubov quasiparticle in the bonding
band and one in the antibonding. 

Figure \ref{fig:BCStp01}(b) shows the real part $\sigma_{c}$ of the $c$-axis
conductivity obtained using the multilayer formula \ref{eq:vdMT} with
$\varepsilon_\mathrm{bl}=\varepsilon_\mathrm{bl/bl}$ and
$\varepsilon_\mathrm{int}=\varepsilon_{\infty}$.  The dominant sharp peaks are
located close to the frequencies of the zero crossings of
$\varepsilon_\mathrm{bl/bl}$.  This can be understood using the fact that for
$|d_\mathrm{bl}\varepsilon_\mathrm{int}|\ll
|d_\mathrm{int}\varepsilon_\mathrm{bl}|$ Eq.~\eqref{eq:vdMT} yields 
\begin{equation}\label{eq:apprepsilon}
\varepsilon(\omega)\approx
\frac{d\varepsilon_\mathrm{int}}{d_\mathrm{int}}
\left(
1-\frac{d_\mathrm{bl}\varepsilon_\mathrm{int}}
{d_\mathrm{int}\varepsilon_\mathrm{bl}}
\right)
\end{equation}
and the expression on the right hand side has poles at the zero crossings of
$\mathrm{Re}\,\varepsilon_{\mathrm{bl}}$.  Physically, the response is similar
to that of a system of thin metallic plates embedded in an insulating matrix,
exhibiting a peak at the plasma frequency of the plates 
(the corresponding effective medium formulas can be found in Ref.~\onlinecite{Aspnes:1982:AJP}).
The~narrow peak at $160\:\mathrm{meV}$ of the NS spectra corresponds to the zero crossing
of $\mathrm{Re}\,\varepsilon_{\mathrm{bl}}$ associated with the interband
transitions, the peak of the SC state spectrum at $60\:\mathrm{meV}$ to the
zero crossing due to the superconducting condensate. 

The VC change the response functions dramatically: the SC-state spectrum of
$\varepsilon_\mathrm{bl/bl}$ shown in Fig.~\ref{fig:BCStp01}(c) displays
neither the superconducting condensate nor the pronounced pair breaking peak.
They are replaced by a broad band centered at $70\:\mathrm{meV}$.  The real
part of $\varepsilon_\mathrm{bl/bl}$ exhibits only two zero crossings 
(instead of the three occurring in the NV case, 
the difference is due to the absence of the condensate).  
The one at lower energies is located in a
region of strong absorption.  As a consequence, the SC-state spectrum of
$\sigma_{c}$ shown in Fig.~\ref{fig:BCStp01}(d) displays only one pronounced
maximum located at the same energy as that of the NS. 

Below we demonstrate,  that the absence of the condensate in
$\varepsilon_\mathrm{bl/bl}$(VC) is a general consequence of the~gauge
invariance.  The current density in the~bilayer region induced
by a homogeneous electric field $\vc E$ oriented along the $c$-axis can
be expressed employing two different gauges of the electromagnetic potentials:\\
(a) $\Delta \varphi=0$, $E_{c}=i\omega A_{c}$.  Here $E_{c}$ is the $c$-axis
component of $\vc E$, $\Delta \varphi$ is the scalar-potential difference
between the two planes,  and $A_{c}$ is the $c$-axis component of the vector
potential;\\ 
(b) $A_{c}=0$, $E_{c}=-\Delta\varphi/d_\mathrm{bl}$.\\ 
Both approaches should lead to the same result.  In the latter case the
expression for the conductivity contains only a regular component
proportional to a current-density correlator.\cite{Schrieffer:1988:Book,Ehrenreich:1966}
The conductivity thus cannot
possess a singular component corresponding to the condensate.  Note that the
above arguments utilizing the two gauges parallel those used when discussing the
response of a homogeneous superconductor to a~longitudinal probe. 

\begin{figure}[tbp]
\includegraphics{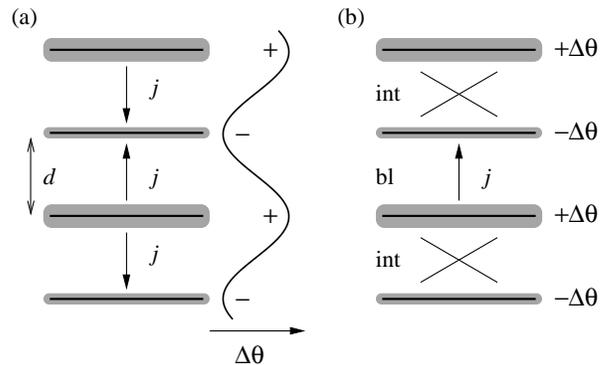}
\caption{(a) Schematic representation of the current density-, 
\mbox{density-,} and phase- pattern 
associated with the Bogolyubov-Anderson mode of a single-layer superconductor 
with $\vc q \parallel c$, $|\vc q|=\pi/d$. 
(b) The same for the collective mode of the bilayer system
discussed in the text.} 
\label{fig:BAscheme}
\end{figure}

The analogy can be further used to understand the nature of the peak (mode)
at $70\:\mathrm{meV}$ in Fig.~\ref{fig:BCStp01}(c).  We recall that
in~homogeneous superconductors a longitudinal electromagnetic field excites
the Bogolyubov-Anderson (B.~A.) mode corresponding to density fluctuations of
the electron system,  associated with a modulation of the phase of the order
parameter.\cite{Bogoljubov:1958:FdP,Anderson:1958:PRB} The energy of the B.~A.~mode 
is proportional to $v_{F}|\vc q|$, where $v_{F}$ is the
Fermi velocity and $\vc q$ the wave-vector.  So far we did not consider the
Coulomb interaction between the carriers, that will shift the mode towards higher
frequencies.  In a single-layer superconductor (one CuO$_{2}$ plane per unit cell), a
longitudinal electromagnetic field with $\vc E \parallel c$ would induce a
B.~A.-like mode with energy proportional to the Fermi velocity along the
$c$-axis $v_{Fz}$, $v_{Fz}\sim t_{\perp}$.  In the present case of the IR
response of a bilayer superconductor the situation is more complicated.  
The electromagnetic wave is transverse with $\vc q \perp c$.  
Nevertheless, it
induces a charge density that is modulated along the $c$-axis.  The modulation
is analogous to the one associated with the B.~A.~mode of a single-layer
superconductor with $\vc q \parallel c$, $|\vc q|=\pi/d$ ($d$ is the interplane
distance).  This is illustrated in Fig.~\ref{fig:BAscheme}.  
The analogy allows us to interpret the mode as an analogue of the B.~A.~mode.
This point of view can be substantiated by comparing the
Eqs.~\eqref{eq:jjvertex}, \eqref{eq:BSE} with those describing the longitudinal
response of a single-layer superconductor.  For $\vc E \parallel c$, $\vc q
\parallel c$, $|\vc q|=\pi/d$, and for the Born-K{\'a}rm{\'an} region
containing only two planes (a rather artificial situation), the latter possess
the same form as the former.  Note that the long-wavelength in-plane
modulation of the electromagnetic wave 
has qualitatively no impact on~the mode. 

\begin{figure}[tbp]
\includegraphics{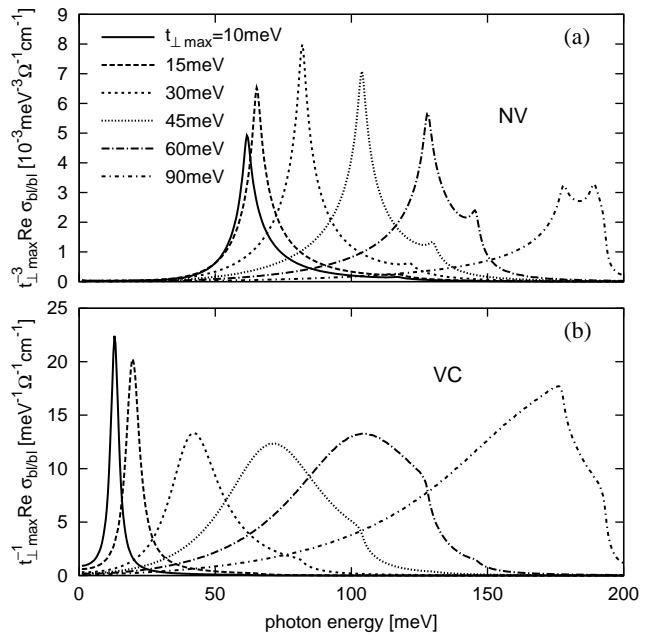}
\caption{Dependence of the real part of the local conductivity
$\sigma_\mathrm{bl/bl}(T=20\:\mathrm{K})$ on the intrabilayer hopping amplitude 
$t_{\perp\mathrm{max}}$ 
calculated with the vertex corrections neglected (a) and included (b). 
In (a), the absorption peak stops at 
$2\Delta_\mathrm{max}=60\:\mathrm{meV}$ 
when decreasing $t_{\perp\mathrm{max}}$. 
In (b), the energy of the peak is proportional to $t_{\perp\mathrm{max}}$.
}
\label{fig:BCStp02}
\end{figure}

Figure \ref{fig:BCStp02} shows the $t_{\perp}$-dependence 
of the intrabilayer conductivity $\sigma_{\mathrm{bl/bl}}$ calculated with the
VC neglected (a) and with the VC included (b).  The frequency of the peak in
(a) is determined by $E_{{\vc k} A}+E_{{\vc k} B}$ which approaches $2\Delta$ for
$t_{\perp}\rightarrow 0$. The energy of the collective mode in (b), however,
does not depend on $\Delta$; instead it is proportional to $t_{\perp}$. This
is consistent with the proposed interpretation of the mode and analogous to
the relation $\omega(\mathrm{B.~A.})\sim v_{F}|\vc q|$. 

A further insight into the origin of the collective mode can be obtained by
using arguments inspired by Anderson's work on gauge invariance and the
Meissner effect.\cite{Anderson:1958:PRB} Anderson explains the difference
between transverse and longitudinal excitations in terms of the complete
second-order phonon-mediated interaction between electrons. The impact of the
relevant interaction terms on longitudinal and transverse excitations
is shown to be fundamentally different. In the longitudinal case, these terms
lead to a restoration of the gauge invariance,  to the absence of the
condensate contribution in the response function,  
and to the presence of a mode at a finite frequency
proportional to the magnitude of the wavevector. 
In the present case, this role is played by 
the interaction terms involving the products of the form 
\begin{equation}
c^\dagger_{{\vc k}B \uparrow}
c^\dagger_{-{\vc k}A \downarrow}
c^{\phantom{\dagger}}_{-{\vc k}'A \downarrow}
c^{\phantom{\dagger}}_{{\vc k}'B \uparrow}
\quad\text{and}\quad
c^\dagger_{{\vc k}A \uparrow}
c^\dagger_{-{\vc k}B \downarrow}
c^{\phantom{\dagger}}_{-{\vc k}'B \downarrow}
c^{\phantom{\dagger}}_{{\vc k}'A \uparrow}\,\,\,.
\end{equation}
They do not belong to the reduced BCS Hamiltonian leading to
Eq.~\eqref{eq:BCSselfE}.  They have, however, a profound impact on the final
states since they provide an attractive interaction between ``elementary
excited states'', \textit{i.e.}, the states created by operators 
$c^\dagger_{{\vc k}A \uparrow}
c^{\phantom{\dagger}}_{{\vc k}B \uparrow}$ and 
$c^\dagger_{{\vc k}A \downarrow}
c^{\phantom{\dagger}}_{{\vc k}B \downarrow}$,
that appear in the expressions for the current density operators on the right
hand side of Eq.~\eqref{eq:jpbl}.  The resulting bound state, \textit{i.e.},
the mode behind the maximum in the spectra, can be thought of as equivalent to
a Cooper bound state of a pair of electrons -- one from the bonding band and
the other from the antibonding band -- superimposed on the BCS ground state of
the two bands.   

It has been shown that the B.~A.~mode can be associated with oscillations of
the phase of the order parameter.  We have checked that for small values of
$t_{\perp}$ the collective mode of our bilayer case can be similarly
associated with oscillations of the relative phase of the two planes. The
pattern of the phase modulation is shown in Fig.~\ref{fig:BAscheme}.  Finally,
the mechanism of the increase of the frequency of the mode when going from the
local conductivity $\sigma_\mathrm{bl/bl}$ to the total conductivity,
involving the Coulomb interaction of the charged planes, is an analogue of the
Anderson-Higgs mechanism. 

\begin{figure}[tbp]
\includegraphics{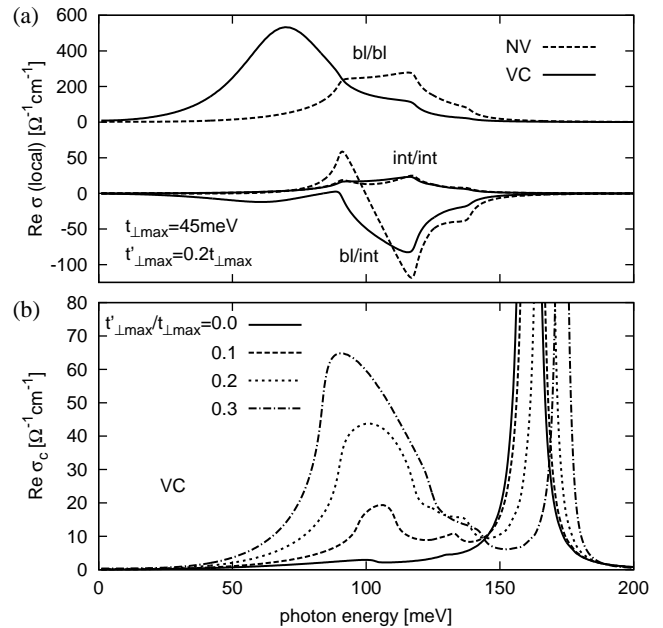}
\caption{
(a) Real parts of the local conductivities 
$\sigma_\mathrm{bl/bl}$, $\sigma_\mathrm{bl/int}$, and $\sigma_\mathrm{int/int}$ 
calculated considering the BCS interaction between the charged quasiparticles 
for $t_{\perp\mathrm{max}}=45\:\mathrm{meV}$, 
$t'_{\perp\mathrm{max}}=0.2t_{\perp\mathrm{max}}$, 
$\Delta_\mathrm{max}=30\:\mathrm{meV}$ and 
$T=20\:\mathrm{K}$.
The spectra of $\mathrm{Re}\,\sigma_\mathrm{bl/int}$ 
and $\mathrm{Re}\,\sigma_\mathrm{int/int}$
are four times magnified.
The dashed (solid) lines correspond to the NV approximation (to the approach
with the VC included).  
(b) Real part of the total $c$-axis conductivity calculated using 
Eq.~\eqref{eq:vdMT2}
for various values of the ratio
$t'_{\perp\mathrm{max}}/t_{\perp\mathrm{max}}$. 
The figure demonstrates the effect of the 
interbilayer hopping.
}
\label{fig:BCStpn01}
\end{figure}

Next we address the more complicated case of $t'_{\perp}\not=0$, where the
theory involves the four local conductivities defined by Eq.~\eqref{eq:Kubo}:
$\sigma_\mathrm{bl/bl}$, $\sigma_\mathrm{bl/int}$, $\sigma_\mathrm{int/bl}$
(that differs from $\sigma_\mathrm{bl/int}$ only by a factor of
$d_\mathrm{bl}/d_\mathrm{int}$),  and $\sigma_\mathrm{int/int}$.  Figure
\ref{fig:BCStpn01}(a) shows the real parts of $\sigma_\mathrm{bl/bl}$,
$\sigma_\mathrm{bl/int}$, and $\sigma_\mathrm{int/int}$ for representative
values of the hopping parameters.  
The dashed (solid) lines correspond to the NV approximation (to the approach
with the VC included).  
In the NV case, all the conductivities display a
pronounced structure around $100\:\mathrm{meV}$: a maximum in
$\mathrm{Re}\,\sigma_\mathrm{bl/bl}$ and
$\mathrm{Re}\,\sigma_\mathrm{int/int}$,
and a wave-like feature in $\mathrm{Re}\,\sigma_\mathrm{bl/int}$.  The VC
lead to drastic changes of $\mathrm{Re}\,\sigma_\mathrm{bl/bl}$.  The maximum
shifts towards lower energies and its spectral weight increases on the account
of the condensate (not shown).  On the other hand, the structures in
$\mathrm{Re}\,\sigma_\mathrm{int/int}$ and $\mathrm{Re}\,\sigma_\mathrm{bl/int}$
remain qualitatively the same and, in particular, they do not shift towards lower
energies. 

\begin{figure}[tbp]
\includegraphics{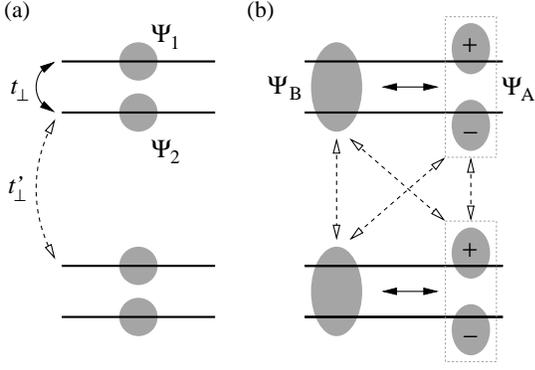}
\caption{Scheme illustrating the differences 
between the local conductivities $\sigma_\mathrm{bl/bl}$ and $\sigma_\mathrm{int/int}$. 
The Wannier-like orbitals of the planes 
are denoted by $\Psi_{1}$ and $\Psi_{2}$, the bonding and antibonding 
orbitals of the individual bilayers by $\Psi_{A}$ and $\Psi_{B}$ respectively. 
The current density operators $\hat{j}^{p}_\mathrm{bl}$ and 
$\hat{j}^{p}_\mathrm{int}$ can be associated with transitions marked
by the solid and the dashed arrows respectively. 
}
\label{fig:Opttrscheme}
\end{figure}

The difference can be understood using Fig.~\ref{fig:Opttrscheme}. 
Part (a) provides a schematic representation of the Wannier-like orbitals of the
planes and of the interplane hopping processes.  For $t'_{\perp}\ll t_{\perp}$, it is
useful to consider bonding and antibonding orbitals of the individual bilayers
shown in (b), $\Psi_{B}=(1/\sqrt{2})[\Psi_{1}+\Psi_{2}]$,
$\Psi_{A}=(1/\sqrt{2})[\Psi_{1}-\Psi_{2}]$.  The local current densities and
conductivities can be discussed and understood in terms of the transitions denoted by the
arrows.  The intra-bilayer current-density operator is connected with
transitions within individual bilayers, marked by the solid arrows.
Note that these transitions create 
{\it two quasiparticles from the same bilayer unit}.  
The inter-bilayer current-density operator is connected with transitions between
adjacent bilayers, marked by the dashed arrows.  These transitions
create {\it two quasiparticles from different units}.  
The important point is that the final states due to the former (latter) transitions 
are strongly (weakly) modified by the VC because {\it the interactions are restricted to
individual bilayers}.  This is the reason, why $\sigma_\mathrm{bl/bl}$
(determined by the matrix element of Eq.~\eqref{eq:jjcorr} 
involving states generated by $\hat{j}^{p}_\mathrm{bl}$, 
\textit{i.e.}, by the former transitions) is strongly influenced  by the VC, 
whereas $\sigma_\mathrm{int/int}$ 
(determined by the matrix element of Eq.~\eqref{eq:jjcorr} 
involving states generated by $\hat{j}^{p}_\mathrm{int}$, 
\textit{i.e.}, by the latter transitions) hardly changes.
The changes of $\sigma_\mathrm{bl/int}$ are more complex
because different final states are involved: one due to the former
transitions and the other due to the latter transitions. 

\begin{figure}[tbp]
\includegraphics[scale=0.96]{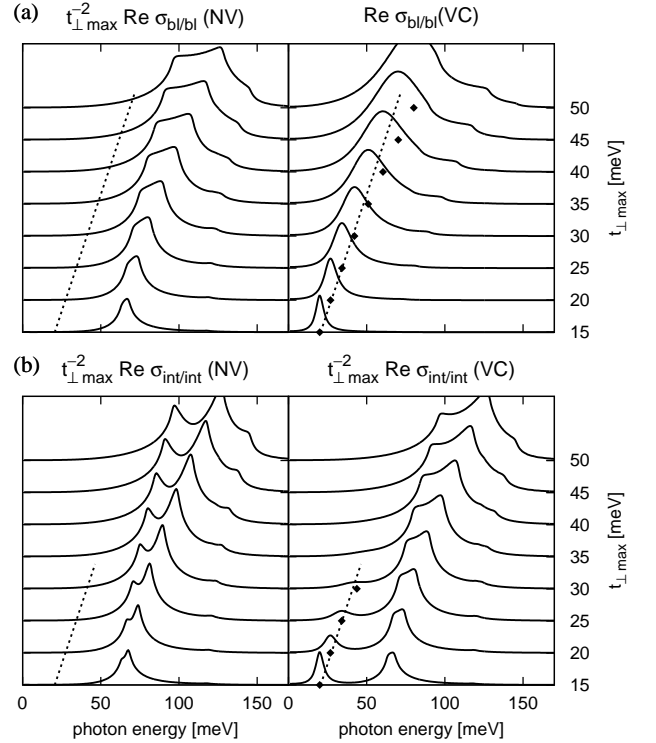}
\caption{
Dependence of $\mathrm{Re}\,\sigma_\mathrm{bl/bl}$ (a) and 
$\mathrm{Re}\,\sigma_\mathrm{int/int}$ (b) on the hopping amplitude $t_{\perp\mathrm{max}}$. 
A constant value of the ratio
$t'_{\perp\mathrm{max}}/t_{\perp\mathrm{max}}$ of $0.2$ has been used.
If the vertex corrections are included, 
the B.~A.-like mode characterized by a linear $t_\perp$-dependence 
appears in the bl/bl component and, for small values of 
$t_{\perp\mathrm{max}}$,  also in the int/int component.
}
\label{fig:BCStpn02}
\end{figure}

The $t_{\perp}$-dependencies of $\mathrm{Re}\,\sigma_\mathrm{bl/bl}$ and
$\mathrm{Re}\,\sigma_\mathrm{int/int}$ are contrasted in
Fig.~\ref{fig:BCStpn02}. 
It can be seen that the VC cause qualitative changes of
$\mathrm{Re}\,\sigma_\mathrm{bl/bl}$, in particular, they lead to the linear
\mbox{$t_{\perp}$-dependence} of the frequency of the maximum.  The spectra of
$\mathrm{Re}\,\sigma_\mathrm{int/int}$, on the other hand, do not change
qualitatively, except for the very small values of $t_{\perp}$, where the
B.~A.~-like mode appears even in the interlayer conductivity. 

The structures of the conductivities $\sigma_\mathrm{int/int}$ and
$\sigma_\mathrm{bl/int}$ give rise -- via the multilayer formula -- to a maximum
in the spectra of $\mathrm{Re}\,\sigma_{c}$ as shown in part (b) of
Fig.~\ref{fig:BCStpn01}.  It can be seen that the magnitude of the peak  is
proportional to $t'^2_{\perp}$.  The maximum of
$\mathrm{Re}\,\sigma_\mathrm{int/int}$ and the related peak of
$\mathrm{Re}\,\sigma_{c}$ can be interpreted simply as an interband
bonding-antibonding pair-breaking (coherence) peak, with the coherence factor 
proportional to the magnitude of the band splitting.  We recall that in
one-band superconductors, the conductivity does not exhibit any coherence peak
around $2\Delta$ due to the fact that electromagnetic absorption belongs
to phenomena governed by case II coherence factors.
\cite{Schrieffer:1988:Book,Tinkham:1996:Book} Here the situation is
different because of the presence of the two bands.  

\subsection{Quasiparticles coupled to spin fluctuations}
\label{secEliashberg}

\begin{figure}[tbp]
\includegraphics{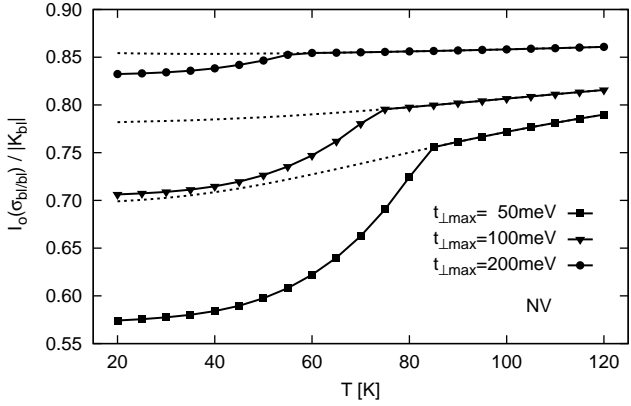}
\caption{Optical weight $I_o=2\hbar^2/\pi e^2\int_{0^+}^\infty 
\mathrm{Re}\,\sigma_\mathrm{bl}(\omega)\,\mathrm{d}\omega$ 
of the local conductivity $\sigma_\mathrm{bl}$ divided by $|K_\mathrm{bl}|$ 
as a function of temperature for various values of the interplane hopping parameter
$t_{\perp}$. Here $K_\mathrm{bl}$ is the effective $c$-axis kinetic energy 
defined by Eq.~\eqref{eq:Efkinen}.  
The results were obtained using Eq.~\eqref{eq:jjsimple} within the spin-fermion model 
with no interbilayer hopping ($t'_{\perp\mathrm{max}}=0$) and the resonant mode 
in the odd interaction channel only, the vertex corrections have been neglected. 
With the inclusion of the vertex corrections, 
$I_o/|K_\mathrm{bl}|$ is always equal to $1$,
both for the normal and the superconducting states.
}
\label{fig:OptSWbl}
\end{figure}

Here we discuss the $c$-axis response obtained by the selfconsistent
computations within the spin-fermion model.  As in the previous section, we
begin with the simpler case of $t'_{\perp}=0$.  Interestingly, the VC play an
important role even in the normal state, an effect that was not addressed at
the BCS level.  In particular, for the NS local conductivity calculated 
with the VC neglected, the restricted conductivity sum rule  
\begin{equation}
I_o=\frac{2\hbar^2}{\pi e^2}\int_{0^+}^\infty \mathrm{Re}\,
\sigma_\mathrm{bl}(\omega)\,\mathrm{d}\omega=-K_\mathrm{bl}
\end{equation}
is strongly (by tens of percent) violated.
This is demonstrated in Fig.~\ref{fig:OptSWbl}. It can be seen that the
deficiency is more serious for smaller values of the band splitting. 
With the VC included, the sum rule is satisfied.  Let us note that the
corresponding violation of the in-plane sum-rule is an order of magnitude
weaker.\cite{Chaloupka:2007:PRB}

\begin{figure}[tbp]
\includegraphics{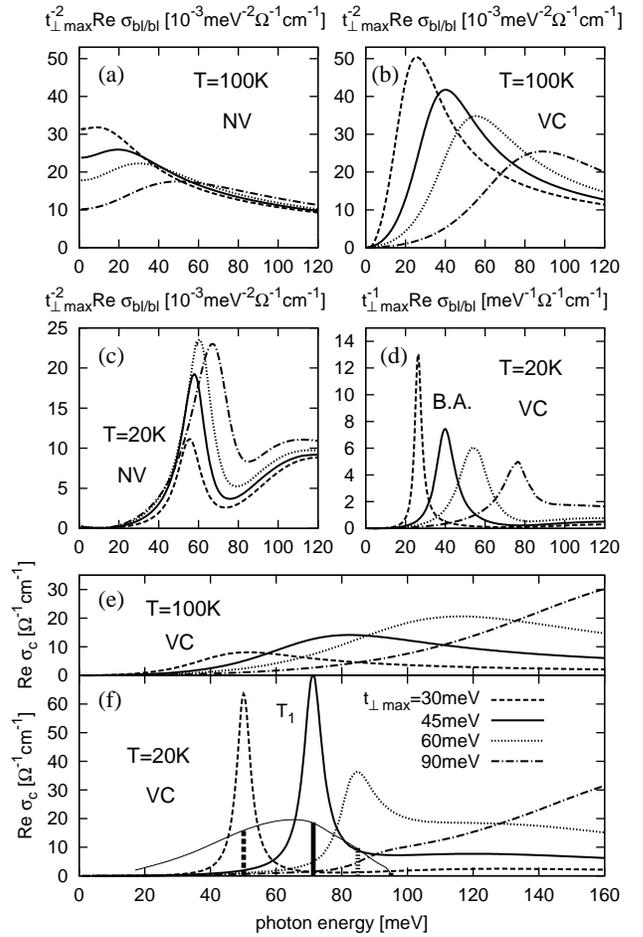}
\caption{
Effect of the vertex corrections on the conductivities. 
Real part of the local conductivity $\sigma_\mathrm{bl/bl}$ [(a)-(d)] and 
of the total $c$-axis conductivity [(e)-(f)] calculated within the spin-fermion model.
The spacing layers have been taken to be insulating, \textit{i.e.}, $t'_{\perp\mathrm{max}}=0$ 
and the neutron resonance has been included in the odd-interaction channel only. 
Results for several values of the hopping parameter $t_{\perp\mathrm{max}}$ are presented. 
The spectra in (a) and (c) [(b), (d), (e), (f)] have been obtained 
for the NV [VC] case.
The thin line in (f) represents the estimated energy dependence 
of the spectral weight of the peak labeled as $T_{1}$. 
For $t_{\perp\mathrm{max}}=30\:\mathrm{meV}$ ($45\:\mathrm{meV}$, $60\:\mathrm{meV}$), 
the estimated value of the spectral weight is 
$4000\:\Omega^{-1}\mathrm{cm}^{-2}$
($4600\:\Omega^{-1}\mathrm{cm}^{-2}$,
$2300\:\Omega^{-1}\mathrm{cm}^{-2}$). 
}
\label{fig:SF1}
\end{figure}

Changes of $\sigma_\mathrm{bl/bl}$ caused by the incorporation of the VC are so
pronounced  that there is not much similarity between the NV and the VC
spectra, see Fig.~\ref{fig:SF1}. 
The NS spectra for the NV case shown in (a) display a broad low-energy absorption band
corresponding to bonding-antibonding interband transitions, that can be
compared with the sharp absorption structure of the NS spectra of the BCS case.  The
broadening with respect to the BCS case is due first to the finite lifetime of
the quasiparticles and second to the presence of a pronounced incoherent
background of the spectral function.  The VC shift the absorption band towards higher
energies, see part (b).  This can be understood in terms of the complete
second-order spin-fluctuation-mediated interaction between the electrons: the
relevant terms can be shown to correspond to a repulsive coupling of the
excited states with one electron in the antibonding band and one hole in the
bonding band. In the total $c$-axis conductivity shown in Fig.~\ref{fig:SF1}(e)
the band is shifted even further and it is very broad. 

The spectra of the SC-state for the NV case as shown in Fig.\ref{fig:SF1}~(c) display a pair breaking
peak at about $2\Delta$, corresponding to a similar feature of the BCS case,
and a continuum with an onset around $80\:\mathrm{meV}$.  In addition,
$\mathrm{Re}\,\sigma_\mathrm{bl/bl}$ also contains the contribution of the
condensate $\sim\delta(\omega)$ (not shown).  The VC transform the spectra in
a similar way as in the nonretarded case, see part (d).  They destroy the
condensate and the pair breaking peak; instead, a sharp maximum (mode) appears,
whose energy is proportional to $t_{\perp}$.  As discussed in
Sec.~\ref{secBCS}, the mode can be interpreted as an analogue of the
B.~A.~mode. In the following we shall call it simply B.~A.~mode.  In the
total $c$-axis conductivity, see Fig.~\ref{fig:SF1}(f), the mode is shifted
towards higher energies by virtue of the Coulomb effects associated with the
charging of the planes, the corresponding peak will be labeled as $T_{1}$.
Let us emphasize that the sharp peak shows up only in the superconducting
state, the presence of a narrow mode in $\chi_\mathrm{SF}$ is not a
sufficient condition for its appearance.  

\begin{figure}[tbp]
\includegraphics{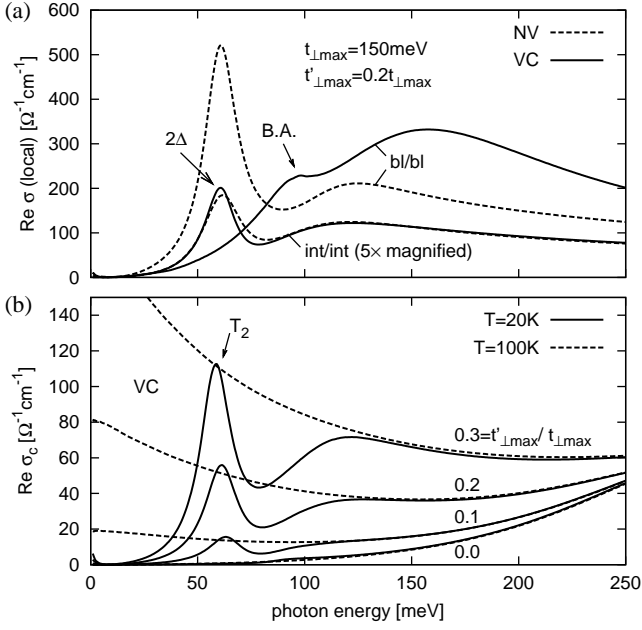}
\caption{ 
Effect of finite interbilayer hopping  ($t'_{\perp\mathrm{max}}>0$) on the conductivities. 
Real parts of the local conductivities $\sigma_\mathrm{bl/bl}$ and $\sigma_\mathrm{int/int}$ (a) 
and of the total $c$-axis conductivity (b) calculated within the spin-fermion model.
The neutron resonance has been distributed equally in both (even and odd) interaction channels.
The dashed (solid) lines in (a) represent the spectra obtained with the VC
neglected (included). The spectra in (b) have been obtained with the VC included, 
the dashed (solid) lines correspond to the normal (superconducting) state, 
results for several values of $t'_{\perp\mathrm{max}}$ are shown.    
}
\label{fig:SF2}
\end{figure}

\begin{figure}[tbp]
\includegraphics[scale=0.98]{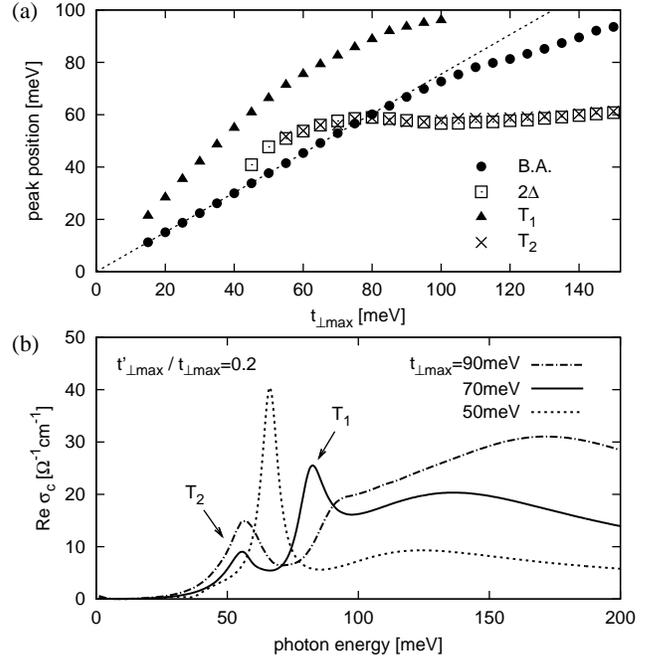}
\caption{
(a) Energies of the B.~A.~peak and of the $2\Delta$-peak 
in the local conductivity $\sigma_\mathrm{bl/bl}$,  
and of the structures $T_{1}$ and $T_{2}$ in the total $c$-axis conductivity 
(for examples, see Figs.~\ref{fig:SF1} and \ref{fig:SF2})
as a function of the intra-bilayer hopping parameter $t_{\perp\mathrm{max}}$. 
A constant value of the ratio $t'_{\perp\mathrm{max}}/t_{\perp\mathrm{max}}$ 
of $0.2$ has been used.
(b) Real part of the total conductivity for values of $t_{\perp\mathrm{max}}$,
where both structures $T_{1}$ and $T_{2}$ are visible. 
As $t_{\perp\mathrm{max}}$ decreases,
the peak $T_{1}$ emerges from the background at
$t_{\perp\mathrm{max}}\approx 100\:\mathrm{meV}$, 
and it quickly becomes the dominant feature.
Eventually, it covers the $T_2$-peak at
$t_{\perp\mathrm{max}}\approx 50\:\mathrm{meV}$.
}
\label{fig:SF3}
\end{figure}

In the remaining part of this subsection, we address the case of nonzero
$t'_{\perp}$.  Similarly as in the BCS case, the impact of the VC on
$\sigma_\mathrm{bl/bl}$ is much stronger than that on the other local
conductivities.  This is illustrated in part (a) of Fig.~\ref{fig:SF2}.
The NV spectra of $\mathrm{Re}\,\sigma_\mathrm{bl/bl}$ exhibit a pair-breaking
peak at about $2\Delta$, similar as in Fig.~\ref{fig:SF1}(c).  The VC
destroy the peak and lead to the formation of the B.~A.~mode.  For the present
value of $t_{\perp}$ of $150\:\mathrm{meV}$, the mode is located in the region
of the continuum [cf.~Fig.~\ref{fig:SF1}(d)] and
thus only gives rise to a weak structure around
$100\:\mathrm{meV}$.  The inter-bilayer conductivity
$\sigma_\mathrm{int/int}$, on the other hand, is almost unaffected by the VC,
retaining the characteristic maximum at about $2\Delta$ (labeled as
$2\Delta$ maximum in the following).  The $t_{\perp}$-dependencies of the
energies of the B.~A.~mode and of the $2\Delta$-peak in the local conductivity
are shown in Fig.~\ref{fig:SF3}(a). 
The energy of the B.~A.~mode can be seen to be proportional to $t_{\perp}$,
except for the region of high values of $t_{\perp}$; that of the $2\Delta$-peak
is approximately $t_{\perp}$-independent. 

The $2\Delta$-maximum appears also in the total $c$-axis conductivity, as
shown in Fig.~\ref{fig:SF2}(b).  The spectral weight of the corresponding peak
(labeled as $T_{2}$ in the following) is proportional to $t'^{2}_{\perp}$.  As
documented in Fig.~\ref{fig:SF3}(b), for low values of $t_{\perp}$,  $T_{1}$
dominates and $T_{2}$ cannot be resolved.  For high values of $t_{\perp}$, on
the other hand, $T_2$ is the most pronounced feature since the B.~A.~mode
merges with the continuum and $T_{1}$ can hardly be resolved.  Both features
can be seen for intermediate values of $t_{\perp}$, \textit{e.g.},
$t_{\perp}=70\:\mathrm{meV}$.  The $t_{\perp}$-dependencies of the energies of
the structures $T_{1}$ and $T_{2}$ in the total $c$-axis conductivity are
given in Fig.~\ref{fig:SF3}(a). 

\subsection{Comparison with experiment and with the theory proposed by Shah and Millis}
\label{secCompofTandE}

The $c$-axis conductivity displays two superconductivity-induced structures
(modes): $P_{1}$ and $P_{2}$ in the experimental data, and
$T_{1}$ and $T_{2}$ in the theoretical spectra.
In what follows, we argue that the
features $P_{1}$ and $P_{2}$ can be attributed to $T_{1}$ and $T_{2}$,
respectively. 

\begin{figure}[tbp]
\includegraphics{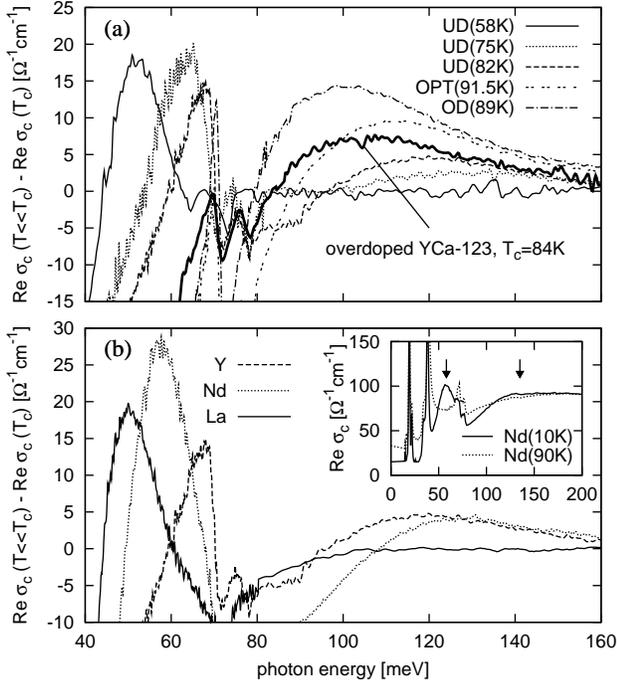}
\caption{
(a)~Doping dependence of the difference 
$\mathrm{Re}\,\sigma_c(T\ll T_c)-\mathrm{Re}\,\sigma_c(T\approx T_c)$
for Y-123. The abbreviations UD, OPT, and OD stand for underdoped, optimum
doped, and overdoped. The values of $p$ are $0.093$, $0.116$, $0.124$,
$0.155$, and $0.194$. Also shown are the data for Y$_{0.86}$Ca$_{0.14}$-123
with $T_c=84\:\mathrm{K}$ and $p=0.176$. Details concerning the samples and the experiment are
given in Refs.~\onlinecite{Yu:2008:PRL}, \onlinecite{Bernhard:2008:JPCS}.
(b)~The spectra of the difference for the sequence
\mbox{$R-123$} (\mbox{$R=Y$}, Nd, and La) with $p\approx 0.12$.
In this sequence, the distance between the \mbox{CuO$_2$} planes
within a bilayer increases. The inset shows the original spectra for Nd-123.
Adapted from Fig.~2 of Ref.~\onlinecite{Yu:2008:PRL}.
}
\label{fig:exper}
\end{figure}
First, we summarize the relevant trends of the structures $P_{1}$ and
$P_{2}$ as observed in the experimental data of bilayer compounds, 
in particular Y-123 and related systems. Some of the trends are
demonstrated in Fig.~\ref{fig:exper}.\\
(E1) The frequency of $P_{1}$ increases with increasing hole concentration
$p$.  \cite{Homes:1993:PRL,Homes:1995:PhysicaC,Schuetzman:1995:PRB,
Munzar:1999:SSC,Grueninger:2000:PRL,Zelezny:2001:PRB,Munzar:2001:PRB,
Dordevic:2004:PRB,Yu:2008:PRL}\\
(E2) The spectral weight (SW) of $P_{1}$ first increases with increasing $p$, 
then saturates for $p\approx 0.12$, and for higher values of $p$,
$P_{1}$ broadens and its SW gradually decreases; for $p>0.15$, $P_{1}$ cannot be
resolved anymore.  \cite{Yu:2008:PRL}\\
(E3) The frequency of $P_{1}$  decreases when going -- for a fixed
doping level -- from Y-123 over Nd-123 to La-123.  \cite{Yu:2008:PRL} In this sequence
of materials, the distance between the closely-spaced planes increases. \\
(E4) In the YPr-123 system, the doping level can be modified either by
changing the oxygen concentration  or by partially replacing Y with Pr.  The
replacement leads to a decrease of~$p$.  By combining the two approaches, it
is possible to obtain various combinations of $p$ and the NS dc-conductivity
along the $c$-axis, $\sigma_\mathrm{dc}$.  For a fixed doping, there is no
pronounced correlation between the SW of $P_{1}$ and $\sigma_\mathrm{dc}$.
\cite{Bernhard:2000:PRB,Dordevic:2004:PRB} \\
(E5) The structure $P_{2}$ can be resolved only for $p \geq 0.10$.
\cite{Yu:2008:PRL} \\ 
(E6) The frequency of $P_{2}$ slowly decreases with increasing~$p$.
\cite{Yu:2008:PRL} \\
(E7) The SW of $P_{2}$ increases with increasing $p$.\cite{Yu:2008:PRL} \\
(E8) In the YCa-123 system, it is possible, similarly as in the case of
YPr-123, to obtain various combinations of $p$ and $\sigma_\mathrm{dc}$.  For
a fixed doping, the SW of $P_{2}$ increases with increasing 
$\sigma_\mathrm{dc}$.  \cite{Bernhard:YCa-123} \\
The distance between the closely-spaced planes can be expected to be
correlated with the strength of the intra-bilayer electronic coupling.  The
observation (E3) thus suggests a relation between $P_{1}$ and the coupling.
Further, the dc-conductivity is likely to reflect the strength of the coupling
through the spacing layer.  The observations (E4) and (E8) thus seem to imply
an independence of $P_{1}$ and a dependence of $P_{2}$ on this coupling. 

Second,  we review the corresponding properties of $T_{1}$ and $T_{2}$ resulting from our
computations.  The labels (T1)--(T8) parallel those used in the summary of experimental
facts. \\ 
(T1) The energy of $T_{1}$ increases with increasing $t_{\perp}$, see
Figs.~\ref{fig:BCStp02} and \ref{fig:SF1}.\\
(T2) The SW of $T_{1}$ first increases with increasing $t_{\perp}$,  then
saturates for $t_{\perp}\approx 50\:\mathrm{meV}$, and for higher values
of $t_{\perp}$,  $T_{1}$ broadens and gradually vanishes, see
Fig.~\ref{fig:SF1}(f).  The broadening is due to the fact that $T_{1}$
reaches the continuum.\\
(T4) For a given $t_{\perp}$, $T_{1}$ does not change significantly with
increasing $t'_{\perp}$ (not shown).  Note that $t'_{\perp}$ determines
$\sigma_\mathrm{dc}$:  $\sigma_\mathrm{dc}$ is approximately proportional to
$t'^{2}_{\perp}$.\\
(T6) The energy of $T_{2}$ is approximately equal to $2\Delta$ for low values
of $t_{\perp}$ and somewhat larger than $2\Delta$ for higher values of
$t_{\perp}$ of the order of $100\:\mathrm{meV}$, see Fig.~\ref{fig:SF3}(a).\\ 
(T7) The SW of $T_{2}$ increases with increasing $t_{\perp}$ (not shown). \\
(T8) The SW of $T_{2}$ also increases with increasing $t'_{\perp}$, see Fig.~\ref{fig:SF2}.
For a given $t_{\perp}$, the SW of $T_{2}$ is approximately proportional to
$t'^{2}_{\perp}$, similarly as $\sigma_\mathrm{dc}$. 

A comparison between the items (E1)--(E4) and (T1)--(T4) suggests that
$P_{1}$ could be attributed to $T_{1}$ \textit{provided that $t_{\perp}$
increases with increasing $p$}.  This crucial assumption is consistent with
results obtained using the bilayer $t-J$ model and the Gutzwiller approximation.
\cite{Mori:2002:PRB,Mori:2006:JofPSJ}  Alternatively, it can also be
justified considering a pseudogap (PG) competing with superconductivity and
the reported $p$- and $\vc k$-dependencies of the magnitude of the PG and of
the coherence peaks due to superconductivity.
\cite{Tanaka:2006:Science,Lee:2007:Nature}  With decreasing $p$, the
magnitude of the PG increases and  the area of the part of the Brillouin zone
dominated by the PG, centered around the antinode, expands.  At the same
time, the area of the part with pronounced Bogolyubov quasiparticles shrinks
towards the BZ diagonal.  The important point is that at the BZ diagonal,
$t_{\perp}$ is probably the smallest.\cite{Andersen:1995:JPCS} The shrinkage
of the area of strong superconducting correlations  might thus lead to a
decrease of an effective $t_{\perp}$ determining the energy of the low-energy
mode. 

The properties (E5)--(E8) are in agreement with attributing $P_{2}$ to $T_{2}$:
The items (E5) and (E7) can be understood in terms of (T7) and (T8), when
combined with the assumption of $t_{\perp}\sim p$ and with the obvious fact
that $t'_{\perp}$ increases with increasing oxygen concentration;  (E6) is
consistent with (T6), when combined with the experimental fact that, around
optimum doping,  $\Delta$ decreases with increasing $p$.  Further,
(E8) can be understood based on (T8). 

\begin{figure}[tbp]
\includegraphics{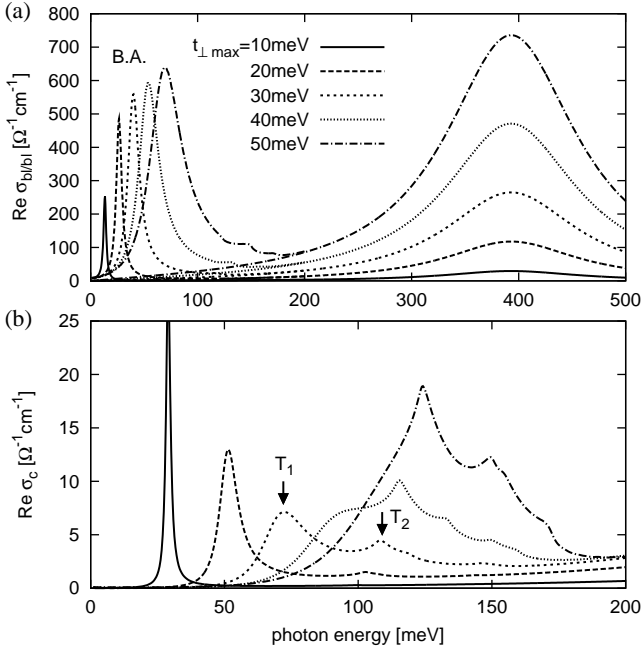}
\caption{Figure illustrating the potential of the model  
to provide the order of the spectral structures consistent with experimental data 
on underdoped Y-123. 
(a) Real part of the local conductivity $\sigma_\mathrm{bl/bl}$ 
consisting of the result of the BCS-approach with $\Delta=50\:\mathrm{meV}$ 
and a broad Lorentzian described in the text. 
(b) Real part of the total \mbox{$c$-axis} conductivity obtained 
using the local conductivity $\sigma_\mathrm{bl/bl}$ shown in (a)
and the other three local conductivities resulting from the BCS-approach. 
A constant value of the ratio $t'_{\perp\mathrm{max}}/t_{\perp\mathrm{max}}$ 
of $0.2$ has been used.
}
\label{fig:BCSandLorentz}
\end{figure}

An obvious problem of the proposed assignment of $T_{1}$ to $P_{1}$ and
$T_{2}$ to $P_{2}$ is that $\omega(T_{1})>\omega(T_{2})$ for the relevant range 
of parameters (see Fig.~\ref{fig:SF3}),  whereas for underdoped materials with $0.10<p<0.15$ the
frequency of $P_{1}$ is lower than that of $P_{2}$.\cite{Yu:2008:PRL}  The
reason for this discrepancy is probably the following: Experimentally,
$\omega(P_{2})\approx 100\:\mathrm{meV}$ which requires $\Delta\approx
50\:\mathrm{meV}$, and such a high value cannot be achieved using the present
selfconsistent theory with reasonable values of input parameters. 
The effective gap may have a contribution due to the pseudogap,
that is not included in the present theory. Within the
BCS approach, $\Delta$ is an input parameter. Unfortunately, the resulting
spectra of $\sigma_c$ for the interesting region of the parameter
space ($\Delta>t_{\perp}$) contain overlapping resonances that cannot be
easily disentangled.  The complication is due to the absence of an incoherent
background in the local conductivities.  Motivated by this observation, we
have supplemented the component $\sigma_\mathrm{bl/bl}$  with a
broad Lorentzian, $-iA/(\omega_{L}^{2}-\omega^{2}-i\omega\Gamma_{L})$,  
$\omega_{L}=0.4\:\mathrm{eV}$,  $\Gamma_{L}=0.15\:\mathrm{eV}$, and $A\sim t_{\perp}^{2}$.  
The $t_{\perp}$-dependence of the spectral-weight parameter 
$A$ is consistent with Eq.~\eqref{eq:jjsimple}.
The results are shown in
Fig.~\ref{fig:BCSandLorentz}. 
It can be seen that the $t_{\perp}$-dependence of the total conductivity 
shown in (b) resembles the doping dependence
of the data \cite{Yu:2008:PRL}, including the interplay of $T_{1}$ and
$T_{2}$: For low values of $t_{\perp}$, the spectra are dominated by $T_{1}$;
for intermediate values of $20-30\:\mathrm{meV}$,  both spectral structures
are present, and with increasing $t_{\perp}$, $T_{1}$ gradually hides in the
continuum.

\begin{figure}[tbp]
\includegraphics{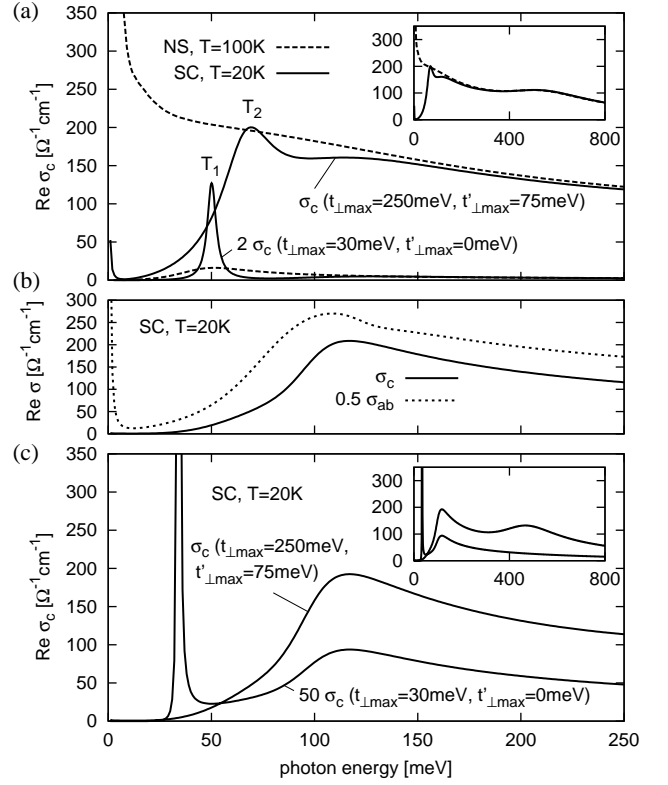}
\caption{Representative examples of the spectra of $\mathrm{Re}\,\sigma_{c}$
compared with the conductivities of a related single-layer system, 
and with the spectra of the same quantity  
obtained using the approximation proposed by Shah and Millis.
(a) Real part of the total $c$-axis conductivity 
calculated using the spin-fermion model 
with $t_{\perp\mathrm{max}}=250\:\mathrm{meV}$, 
$t'_{\perp\mathrm{max}}=75\:\mathrm{meV}$, and 
with $t_{\perp\mathrm{max}}=30\:\mathrm{meV}$, 
$t'_{\perp\mathrm{max}}=0\:\mathrm{meV}$. 
The solid and the dashed lines correspond to the superconducting 
and the normal state, respectively. 
The spectra for 
$t_{\perp\mathrm{max}}=250\:\mathrm{meV}$ and 
$t'_{\perp\mathrm{max}}=75\:\mathrm{meV}$ 
on an extended scale are shown in the inset. 
(b) Real parts of the in-plane conductivity and of the $c$-axis conductivity 
for a single-layer superconductor with the plane spacing  
of $d_\mathrm{bl}+d_\mathrm{int}$ 
and with the $c$-axis hopping parameter of the form of Eq.~\eqref{eq:tperp}
and the maximum of $75\:\mathrm{meV}$. 
(c) The same as in (a) but using the approach proposed 
by Shah and Millis, where the nondiagonal components of the conductivity 
$\sigma_\mathrm{bl/int}$ and $\sigma_\mathrm{int/bl}$ are neglected,  
and the diagonal component $\sigma_\mathrm{bl/bl}$ ($\sigma_\mathrm{int/int}$)
is approximated  by the conductivity 
of the single-layer superconductor with the hopping parameter equal to 
$t_{\perp}$ ($t'_{\perp}$). 
Only the superconducting state spectra are shown. 
}
\label{fig:sigmacover}
\end{figure}

The discrepancy in the order of the spectral structures does not occur for
optimally doped and overdoped Y-123 samples, where $P_{1}$ is buried in the
continuum part of the spectra located above the maximum $P_{2}$.  The
experimental spectra (see \textit{e.g.} Fig.~1 of
Ref.~\onlinecite{Grueninger:2000:PRL}) are similar to the calculated ones
corresponding to the values of $t_{\perp}$ of a few hundreds of meV.  As an
example, we show in Fig.~\ref{fig:sigmacover}(a)
the NS and SC-state spectra of $\mathrm{Re}\,\sigma_{c}$ corresponding 
to $t_{\perp\mathrm{max}}=250\:\mathrm{meV}$ and
$t'_{\perp\mathrm{max}}=0.3t_{\perp\mathrm{max}}$.  
The maximum $T_{2}$ at about $70\:\mathrm{meV}$ ($600\:\mathrm{cm}^{-1}$) 
probably corresponds to $P_{2}$ occurring 
at a slightly higher frequency in the experimental spectra. 

Next we compare our calculated spectra of $\mathrm{Re}\,\sigma_{c}$ with the
conductivities of a related single-layer superconductor and with those
computed along the lines of the theory proposed by Shah and Millis
(SM).\cite{Shah:2001:PRB} Figure~\ref{fig:sigmacover}(a) shows, besides the
spectra of $\mathrm{Re}\,\sigma_{c}$
($t_{\perp\mathrm{max}}=250\:\mathrm{meV}$,
$t_{\perp\mathrm{max}}=75\:\mathrm{meV}$) discussed above, those corresponding
to $t_{\perp\mathrm{max}}=30\:\mathrm{meV}$,
$t_{\perp\mathrm{max}}=0\:\mathrm{meV}$.  The former (the latter) represent
the case of strong (weak) intra-bilayer coupling.  The corresponding
renormalized values of the (normal-state) bilayer splitting are ca $80$ and
$10\:\mathrm{meV}$.  Note that the former value is close to that of Bi-2212 as
obtained by photoemission experiments
\cite{Feng:2001:PRL,Chuang:2001:PRL,Kordyuk:2002:PRL,Borisenko:2002:PRB}. As
discussed in the context of Figs.~\ref{fig:SF2} and \ref{fig:SF3} only one
peak is present in the superconducting state spectra: $T_{2}$ in the former
case and $T_{1}$ in the latter.  It is instructive to compare the SC-state
conductivities with the solid line of Fig.~\ref{fig:sigmacover}(b)
representing the $c$-axis conductivity of a model single-layer superconductor
described in the caption.  This allows one to identify the features specific
to the bilayer compounds: (\textit{i})~the peak $T_{1}$ (for small values of
$t_{\perp}$), (\textit{ii})~the peak $T_{2}$ (for large values of $t_{\perp}$
and $t'_{\perp}\not=0$), and (\textit{iii})~a hump in the mid-infrared (for
large values of $t_{\perp}$).  For $t_{\perp\mathrm{max}}=250\:\mathrm{meV}$,
the hump is centered at $500\:\mathrm{meV}$.  It has the same origin as the
$160\:\mathrm{meV}$ maximum in Fig.~\ref{fig:BCStp01} and it can be attributed
to the (upper) plasma mode of the bilayer unit, see the discussion following
Eq.~\eqref{eq:apprepsilon}.  Both $T_{1}$ and $T_{2}$ would appear for
intermediate values of $t_{\perp}$, as discussed in the context of
Fig.~\ref{fig:SF3} and Fig.~\ref{fig:BCSandLorentz}.  The broad band centered
around $120\:\mathrm{meV}$ in Fig.~\ref{fig:sigmacover}(a) appears also in the
conductivity of the single-layer superconductor and is thus not specific to
the bilayer compounds. 

Also shown in Fig.~\ref{fig:sigmacover}(b) is the in-plane conductivity of the
single-layer superconductor. It can be seen that the shapes of the two
conductivities are fairly similar.  The main differences are: (a) for
$T=20\:\mathrm{K}$, a narrow Drude term can be resolved only in
$\mathrm{Re}\,\sigma_{ab}$, and (b) the onset of $\mathrm{Re}\,\sigma_{c}$ is
more gradual than that of $\mathrm{Re}\,\sigma_{ab}$.  Both (a) and (b) are
due to the difference between the matrix element of $\sigma_{ab}$,
\textit{i.e.}, the in-plane quasiparticle velocity,  and $t_{\perp}$.  Both
$\sigma_{ab}$ and $\sigma_{c}$ exhibit a maximum around $110\:\mathrm{meV}$
and both decrease with increasing energy above this maximum. The origin of the
maximum has been addressed in Refs.~\onlinecite{Casek:2005:PRB},
\onlinecite{Chaloupka:2007:PRB}.  The calculated $c$-axis conductivity of the
single-layer superconductor is qualitatively similar to the measured
conductivity of \mbox{La$_{2-x}$Sr$_x$CuO$_4$} reported in
Ref.~\onlinecite{Kuzmenko:2003:PRL}. 

Figure \ref{fig:sigmacover}(c) shows the $c$-axis conductivities of the same
bilayer systems as in (a), but calculated using the approach proposed by SM,
where the nondiagonal components of the conductivity $\sigma_\mathrm{bl/int}$
and $\sigma_\mathrm{int/bl}$ are neglected,  and the diagonal component
$\sigma_\mathrm{bl/bl}$ ($\sigma_\mathrm{int/int}$) is approximated  by the
conductivity of the single-layer superconductor with the hopping parameter
equal to $t_{\perp}$ ($t'_{\perp}$).  When calculating the latter
conductivities, SM further replace the $k_{z}$-dependent Green's function with
a $k_{z}$-independent one of a two-dimensional model.  Instead of using this
approximation, we have obtained the conductivities of the model single-layer
superconductors by scaling the one shown in Fig.~\ref{fig:sigmacover}(b).  We
have checked that for the present values of the parameters, the results of the
two approaches are almost the same.  By comparing panels (a) and (c) of
Fig.~\ref{fig:sigmacover}, we easily identify differences between the SM
theory that does not involve the bilayer splitting and our improved approach,
that does.  For small values of $t_{\perp}$, the SM theory provides a peak
inside the gap, corresponding to a bilayer plasmon, similar to the transverse
Josephson plasmon of the phenomenological  model.\cite{Marel:1996:CzJP} The
$T_{1}$ peak of our theory is located at a slightly higher energy and its
interpretation is different. What it has in common with the transverse
Josephson plasmon is that both are associated with oscillations of the
relative phase of the two closely-spaced planes.  For high values of
$t_{\perp}$, our approach yields the pair-breaking peak $T_{2}$ absent at the
SM level.  At high energies (above $100\:\mathrm{meV}$), the results of the
two approaches are similar. 

Interestingly, the NS spectra corresponding to
$t_{\perp\mathrm{max}}=250\:\mathrm{meV}$ shown in
Fig.~\ref{fig:sigmacover}(a) do not display any clear signature of the bilayer
splitting, consistent with experimental data.  The contribution of the
bonding-antibonding transitions is hidden in the mid-infrared region. This is
because the renormalization of the quasiparticles leads to a very broad
absorption band [see Fig.~\ref{fig:SF1}(a)] shifted towards the mid-infrared
both by the the vertex corrections and by the Coulomb effects [see
Fig.~\ref{fig:SF1}(b) and (e)].  

Experiments reveal an increase of the optical spectral weight in the
far-infrared below $T_{c}$, with a possible interpretation in terms of a
decrease of the $c$-axis kinetic energy associated with the superconducting
transition.\cite{Basov:1999:Science} The picture resulting from our
calculations, restricted to the case of small values of
$t_{\perp\mathrm{max}}$ and $t'_{\perp\mathrm{max}}=0$, is the following:
Below $T_{c}$, the peak $T_{1}$ forms, gaining spectral weight from a broad
interval of energies, and the spectral weight at low energies increases.  The
total spectral weight, proportional to the negatively taken effective kinetic
energy $K_\mathrm{bl}$, decreases, provided that $\chi_\mathrm{SF}$ (below
$T_{c}$)= $\chi_\mathrm{SF}$ (above $T_{c}$).  However, changes of
$\chi_\mathrm{SF}$ upon entering the superconducting state, in particular the
formation of the resonance mode, can lead to a slight increase of the total
spectral weight and the corresponding decrease of $K_\mathrm{bl}$, the
mechanism being connected to that outlined in
Ref.~\onlinecite{Norman:2000:PRB}.  The issue is fairly complex and will be
addressed in a separate publication.

\section{SUMMARY AND CONCLUSIONS}
\label{secSM}

We have constructed a realistic microscopic model of the $c$-axis infrared
response of bilayer cuprate superconductors allowing us to interpret the
superconductivity-induced modes occurring in the experimental data.  

For the simpler case of insulating spacing layers, the local conductivity of
the intrabilayer region does not possess a condensate contribution in the
superconducting state [$\delta(\omega)$ in $\mathrm{Re}\,\sigma$] as assumed
within the phenomenological Josephson superlattice model.  Instead, it
displays a collective mode at a finite frequency that is proportional to the
interplane hopping amplitude $t_{\perp}$.  This has been shown to be a
consequence of the gauge invariance. The nature of the mode is similar to that
of the Bogolyubov-Anderson mode that participates in the longitudinal response
of a homogeneous superconductor. It is associated with charge oscillations
between the planes and, for small values of $t_{\perp}$, also with
oscillations of the relative phase of the two planes.  This physical picture
is fairly similar to that of the transverse plasmon of the Josephson
superlattice model.  In the total $c$-axis conductivity the mode is shifted
towards higher energies by the interplane Coulomb interaction.

A nonzero amplitude of the hopping through the spacing layer implies a finite
conductivity of this layer. This local conductivity exhibits a peak at a
frequency slightly higher than $2\Delta_{\mathrm{max}}$, that can be
interpreted as a pair breaking peak.  The simple picture is such that two
Bogolyubov quasiparticles are involved: one from the bonding band and the
other from the antibonding. The reason, why the peak appears in the
inter-bilayer conductivity and not in the intra-bilayer one, is the following.
Excited states behind the former (latter) conductivity involve quasiparticles
from different bilayers (from the same bilayer). Only the latter are thus
strongly modified by including the final-state interactions, that are
restricted to individual bilayers. The peak permeates into the total $c$-axis
conductivity.

A series of arguments has been presented assigning the collective mode to the
low energy superconductivity-induced mode of the experimental data
(interpreted previously in terms of the Josephson superlattice model) and the
pair-breaking maximum to the superconductivity-induced peak centered around
$1000\:\mathrm{cm}^{-1}$.  The arguments concern the doping dependence of the
frequencies and the spectral weights of the peaks and the impact of various
substitutions. A crucial assumption,  connecting the theory and the
experiment, is that the effective $t_{\perp}$ decreases with decreasing
doping.  The trends of the underdoped regime, in particular, the appearance of
the collective mode below $T_{c}$, the increase of its frequency with
increasing doping and its disappearance below optimum doping  can all be
reasonably reproduced and understood using this assumption.  Admittedly, the
values of $t_{\perp}$ of a few tens of meV needed to fit the data are smaller
than those deduced from photoemission experiments.  The main features of the
data of optimally doped Y-123 can be reasonably reproduced with
$t_{\perp\mathrm{max}}$ of $250\,\mathrm{meV}$, which corresponds to the
maximum distance between the renormalized bands of ca $80\:\mathrm{meV}$.  

\section*{ACKNOWLEDGMENTS}

This work was supported by the Ministry of Education of Czech Republic
(MSM0021622410) and the Schweizerische Nationalfonds (SNF) by grant
200020-119784. In an early stage of this research, during a stay at MPI
Stuttgart, D.~M. was supported by the AvH Foundation.  J.~Ch.~thanks B.~Keimer
and G.~Khaliullin for their hospitality during a stay at MPI Stuttgart, where
a part of this work was performed.  We gratefully acknowledge helpful
discussions with Li~Yu, A.~Dubroka, J.~Huml\'{\i}\v{c}ek, B.~Keimer, R.~Zeyher
and J.~Va\v{s}\'{a}tko.

\appendix

\section{BCS level of the theory}
\label{appBCS}

In this appendix, we show several results obtained at the BCS level, where
extensive analytical simplifications can be made. For the sake of brevity, we
restrict ourselves to the case of $\tppk=0$. We employ the BCS interaction
of \mbox{$d$-wave} symmetry which acts in the individual CuO$_2$ planes 
(labeled as $1$ and $2$) within a bilayer unit. The corresponding Hamiltonian 
reads
\begin{multline}\label{eq:BCSham1}
\hat{H}_\mathrm{BCS} = 
\sum_{\vc k \sigma} \varepsilon_{B\vc k}
 c_{B\vc k\sigma}^\dagger c_{B\vc k\sigma}^{\phantom{\dagger}}
+
\sum_{\vc k \sigma} \varepsilon_{A\vc k}
 c_{A\vc k\sigma}^\dagger c_{A\vc k\sigma}^{\phantom{\dagger}}
+ \\ +
\sum_{\vc k \vc k', n\in\{1,2\}} 2V_{\vc k\vc k'}
c_{n\vc k\uparrow}^\dagger c_{n,-\vc k\downarrow}^\dagger
c_{n,-\vc k'\downarrow}^{\phantom{\dagger}}
c_{n\vc k'\uparrow}^{\phantom{\dagger}} \;,
\end{multline}
where $V_{\vc k\vc k'}$ is introduced in the main text. The factor of $2$ is
for later convenience. The interaction term can be written as
\begin{multline}\label{eq:BCSham2}
\sum_{\vc k\vc k'} V_{\vc k\vc k'} ( BBBB
+BBAA+AABB+AAAA + \\  BABA + BAAB + ABAB + ABBA)\;.
\end{multline}
where $BBBB$, \textit{e.g.}, stands for 
$c_{B\vc k\uparrow}^\dagger c_{B,-\vc k\downarrow}^\dagger
c_{B,-\vc k'\downarrow}^{\phantom{\dagger}}
c_{B\vc k'\uparrow}^{\phantom{\dagger}}$.
The first four terms in Eq.~\eqref{eq:BCSham2} provide the pairing
interaction, the other four terms play an important role in the vertex
corrections. The above pairing interaction is equally distributed among the
two symmetry channels and produces the same gap in both bands, hence
$\Sigma_{A/B}(\vc k)=-\Delta_{\vc k}\tau_1$ with $\Delta_{\vc k}$ determined
by Eq.~\eqref{eq:BCSselfE}. 

Because the selfenergy depends on $\vc k$ only, we can sum over the Matsubara
frequencies explicitly. The evaluation of the NV response function given by
Eq.~\eqref{eq:jjsimple} leads to
\begin{multline}\label{eq:BCSrespNV}
\Pi_\mathrm{bl-bl}^{\mathrm{NV}(1)}(\vc q=0,\hbar\omega) = 
-\frac{e^2}{\hbar^2}\frac{d_\mathrm{bl}}{Na^2} \sum_{\vc k}
t_{\perp{\vc k}}^2 \times \\
\Biggl\{
l_1 [1-n_F(E_{A\vc k})-n_F(E_{B\vc k})] \left( 
\frac1{\hbar\omega+E^+_{\vc k}+i\delta}-
\frac1{\hbar\omega-E^+_{\vc k}+i\delta}
\right) \\
+ l_2 [n_F(E_{A\vc k})-n_F(E_{B\vc k})] \left( 
\frac1{\hbar\omega+E^-_{\vc k}+i\delta}-
\frac1{\hbar\omega-E^-_{\vc k}+i\delta}
\right) \Biggr\} \;,
\end{multline}
where $E_{\vc k}^\pm = E_{A\vc k}\pm E_{B\vc k}$
and $l_{1/2}$ are the coherence factors
\begin{equation}
l_{1/2} = 
\frac{\varepsilon_{A\vc k}\varepsilon_{B\vc k}
     +\Delta_{A\vc k}\Delta_{B\vc k}}{E_{A\vc k}\,E_{B\vc k}}\mp 1 \;.
\end{equation}
Note that in the limit of $t_{\perp\vc k}\rightarrow 0$, the factors $l_1$ as
well as $n_F(E_{A\vc k})-n_F(E_{B\vc k})$ vanish and $\Pi_\mathrm{bl-bl}$
becomes zero. For the same reason, there is no intraband contribution
\eqref{eq:jjintra} in the BCS case.

The renormalized vertices $\Gamma_{AB}$ and $\Gamma_{BA}$ satisfy the
Bethe-Salpeter equations similar to Eq.~\eqref{eq:BSE}, now containing 
the BCS interaction:
\begin{multline}\label{eq:BSEBCS}
\Gamma_{AB}(\vc k,iE,i\hbar\nu) = t_{\perp\vc k}\tau_0 
- \frac{k_BT}N \sum_{\vc k',iE'} V_{\vc k\vc k'} \times \\
\tau_3\Bigl[
\mathcal{G}_B(\vc k',iE')\,\Gamma_{AB}(\vc k',iE',i\hbar\nu)
\mathcal{G}_A(\vc k',iE'+i\hbar\nu) +\\+ \mathcal{G}_A(\vc k',iE')
\,\Gamma_{BA}(\vc k',iE',i\hbar\nu)\mathcal{G}_B(\vc k',iE'+i\hbar\nu)
\Bigr]\tau_3 \;.
\end{multline}
Since the usual BCS interaction couples with $\tau_3$, not with
$\tau_0$ like the spin-fluctuations, the above equation contains
the additional matrices $\tau_3$. The corresponding equation for the
vertex $\Gamma_{BA}$ differs from Eq.~\eqref{eq:BSEBCS} in the 
sign of the $t_{\perp\vc k}\tau_0$ term only.
By inserting the form of $V_{\vc k\vc k'}$ explicitly, one finds, that the 
vertices can be cast to:
$\Gamma_{AB}(\vc k,i\hbar\nu)= 
 t_{\perp\vc k} \tau_0 +\lambda w_{\vc k} C(i\hbar\nu)$
and
$\Gamma_{BA}(\vc k,i\hbar\nu)= 
 -t_{\perp\vc k} \tau_0 +\lambda w_{\vc k} C(i\hbar\nu)$.
The quantity $C(i\hbar\nu)$ is a $2\times2$ matrix, independent of $iE$
because of the non-retarded BCS interaction and independent of $\vc k$ because
of the separable form of $V_{\vc k\vc k'}$. We express it as a linear
combination of the Pauli matrices:
$C(i\hbar\nu) = \sum_{\alpha=0}^3 \tau_\alpha C_\alpha(i\hbar\nu)$.
The Bethe-Salpeter equation \eqref{eq:BSEBCS} can be then converted to a linear
system of equations for the coefficients $C_\alpha(i\hbar\nu)$
\begin{equation}\label{eq:BCSlinsysVC}
\eta_\alpha C_\alpha(i\hbar\nu) 
-\sum_{\beta} \Lambda_{\alpha\beta}(i\hbar\nu)\, C_\beta(i\hbar\nu) = 
\Phi_{\alpha 0}(i\hbar\nu) \;,\; \alpha,\beta=0,1,2,3
\end{equation}
with $\eta_\alpha=+1$ ($\alpha=0,3$), $\eta_\alpha=-1$ ($\alpha=1,2$).
The coefficients of the system are given by
\begin{multline}\label{eq:BCSsyscfP}
\sum_\alpha \tau_\alpha\Phi_{\alpha\beta}=
\lambda\frac{k_BT}N\sum_{\vc k,iE} w_{\vc k} t_{\perp\vc k} 
\bigl[\mathcal{G}_B(\vc k,iE)\,\tau_\beta\,\mathcal{G}_A(\vc k,iE+i\hbar\nu)
\\
-\mathcal{G}_A(\vc k,iE)\,\tau_\beta\,\mathcal{G}_B(\vc k,iE+i\hbar\nu)\bigr] 
\end{multline}
and
\begin{multline}\label{eq:BCSsyscfL}
\sum_\alpha \tau_\alpha\Lambda_{\alpha\beta}=
\lambda\frac{k_BT}N\sum_{\vc k,iE} w_{\vc k}^2
\bigl[\mathcal{G}_B(\vc k,iE)\,\tau_\beta\,\mathcal{G}_A(\vc k,iE+i\hbar\nu)
\\
+\mathcal{G}_A(\vc k,iE)\,\tau_\beta\,\mathcal{G}_B(\vc k,iE+i\hbar\nu)\bigr]
\;.
\end{multline}
Finally, to get the response function, we insert the renormalized vertices
into \eqref{eq:jjvertex} and with the help of
$\mathrm{Tr}\,\tau_\alpha=2\delta_{\alpha 0}$ obtain
\begin{equation}\label{eq:BCSrespVC}
\Pi_\mathrm{bl-bl}^\mathrm{VC}=\Pi_\mathrm{bl-bl}^\mathrm{NV} 
-\frac{2e^2 d_\mathrm{bl}}{\hbar^2a^2} \sum_{\beta} \Phi_{0\beta} C_\beta \;.
\end{equation}
All the above equations can be analytically continued to the real axis
explicitly. For each frequency required, we have to evaluate the coefficients 
$\Phi_{\alpha\beta}$ and $\Lambda_{\alpha\beta}$, solve the linear system 
\eqref{eq:BCSlinsysVC} and find the response function \eqref{eq:BCSrespVC}.
Thanks to the simple form of the selfenergy, the Matsubara summations can be
again performed analytically, as in Eq.~\eqref{eq:BCSrespNV}, but lead to more
cumbersome expressions due to the additional $\tau_\beta$ matrix in
Eqs.~\ref{eq:BCSsyscfP} and \ref{eq:BCSsyscfL}.


\begin{thebibliography}{99}

\bibitem{Homes:1995:PhysicaC}
C.C.~Homes, T.~Timusk, R.~Liang, D.A.~Bonn, and W.N.~Hardy,
Physica C \textbf{254}, 265 (1995).

\bibitem{Schuetzman:1995:PRB}
J.~Sch\"{u}tzmann, S.~Tajima, S.~Miyamoto, M.~Sato, and R.~Hauff,
Phys. Rev. B \textbf{52}, 13665 (1995).

\bibitem{Bernhard:1999:PRB}
C.~Bernhard, D.~Munzar, A.~Wittlin, W.~K\"{o}nig, A.~Golnik, C.T.~Lin,
M.~Kl\"{a}ser, Th.~Wolf, G.~M\"{u}ller-Vogt, and M.~Cardona,
Phys. Rev. B \textbf{59}, R6631 (1999).

\bibitem{Tamasaku:1992:PRL}
K.~Tamasaku, Y.~Nakamura, and S.~Uchida,
Phys. Rev. Lett. \textbf{69}, 1455 (1992).

\bibitem{Homes:1993:PRL}
C.C.~Homes, T.~Timusk, R.~Liang, D.A.~Bonn, and W.N.~Hardy,
Phys. Rev. Lett. \textbf{71}, 1645 (1993).

\bibitem{Basov:2005:RMP}
D.N.~Basov and T.~Timusk,
Rev. Mod. Phys. \textbf{77}, 721 (2005).

\bibitem{Chakravarty:2004:Nature}
S.~Chakravarty, H.Y.~Kee, K.~Volker, Nature \textbf{428}, 53 (2004).

\bibitem{Munzar:2003:PRB}
D.~Munzar, T.~Holden, C.~Bernhard, Phys. Rev. B \textbf{67}, 020501(R) (2003).

\bibitem{Andersen:1995:JPCS}
O.K.~Andersen, A.I.~Liechtenstein, O.~Jepsen, and F.~Paulsen,
J. Phys. Chem. Solids \textbf{56}, 1573 (1995).

\bibitem{Mazin:1992:PRB}
I.I.~Mazin, S.N.~Rashkeev, A.I.~Liechtenstein, and O.K.~Andersen,
Phys. Rev. B \textbf{46}, 11232 (1992).

\bibitem{Munzar:1999:SSC}
D.~Munzar, C.~Bernhard, A.~Golnik, J.~Huml\'{\i}\v{c}ek,
M.~Cardona, Solid State Commun. \textbf{112}, 365 (1999).

\bibitem{Grueninger:2000:PRL}
M.~Gr{\"u}ninger, D.~van~der~Marel, A.A.~Tsvetkov, and A.~Erb,
Phys. Rev. Lett. \textbf{84}, 1575 (2000).

\bibitem{Marel:1996:CzJP}
D.~van der Marel, A.~Tsvetkov, Czech J. Phys \textbf{46}, 3165 (1996).

\bibitem{Shah:2001:PRB}
N.~Shah, A.~J.~Millis,
Phys.~Rev.~B \textbf{65}, 024506 (2001).

\bibitem{Borisenko:2006:PRL}
S.V.~Borisenko, A.A.~Kordyuk, V.~Zabolotnyy, J.~Geck, D.~Inosov,
A.~Koitzsch, J.~Fink, M.~Knupfer, B.~B\"{u}chner, V.~Hinkov, C.T.~Lin,
B.~Keimer, T.~Wolf, S.G.~Chiuzb\u{a}ian, L.~Patthey, and R.~Follath,
Phys. Rev. Lett. \textbf{96}, 117004 (2006).

\bibitem{Feng:2001:PRL}
D.L.~Feng, N.P.~Armitage, D.H.~Lu, A.~Damascelli, J.P.~Hu,
P.~Bogdanov, A.~Lanzara, F.~Ronning, K.M.~Shen, H.~Eisaki,
C.~Kim, J.-i.~Shimoyama, K.~Kishio, and Z.-X.~Shen,
Phys. Rev. Lett. \textbf{86}, 5550 (2001).

\bibitem{Chuang:2001:PRL}
Y.-D.~Chuang, A.D.~Gromko, A.~Fedorov, Y.~Aiura, K.~Oka,
Yoichi~Ando, H.~Eisaki, S.I.~Uchida, and D.S.~Dessau,
Phys. Rev. Lett. \textbf{87}, 117002 (2001).

\bibitem{Kordyuk:2002:PRL}
A.A.~Kordyuk, S.V.~Borisenko, T.K.~Kim, K.A.~Nenkov, M.~Knupfer,
J.~Fink, M.S.~Golden, H.~Berger, and R.~Follath,
Phys.  Rev. Lett. \textbf{89}, 077003 (2002).

\bibitem{Borisenko:2002:PRB}
S.V.~Borisenko, A.A.~Kordyuk, T.K.~Kim, S.~Legner, K.A.~Nenkov,
M.~Knupfer, M.S.~Golden, J.~Fink, H.~Berger, and R.~Follath,
Phys. Rev. B \textbf{66}, 140509(R) (2002).

\bibitem{Yu:2008:PRL}
Li Yu, D.~Munzar, A.V.~Boris, P.~Yordanov, J.~Chaloupka, T.~Wolf,
C.T.~Lin, B.~Keimer, C.~Bernhard, Phys. Rev. Lett. \textbf{100}, 177004 (2008).

\bibitem{Dordevic:2004:PRB}
S.V.~Dordevic, E.J.~Singley, J.H.~Kim, M.B.~Maple, Seiki Komiya,
S.~Ono, Yoichi Ando, T.~R\~{o}\~{o}m, Ruxing Liang, D.A.~Bonn, W.N.~Hardy,
J.P.~Carbotte, C.C.~Homes, M.~Strongin, D.N.~Basov,
Phys. Rev. B \textbf{69}, 094511 (2004).

\bibitem{Chaloupka:2007:PRB}
J.~Chaloupka, D.~Munzar, Phys. Rev. B \textbf{76}, 214502 (2007).

\bibitem{Liechtenstein:1996:PRB}
A.I.~Liechtenstein, O.~Gunnarsson, O.K.~Andersen, and R.M.~Martin,
Phys. Rev. B \textbf{54}, 12505 (1996).

\bibitem{Eschrig:2002:PRL}
M.~Eschrig, M.R.~Norman, Phys. Rev. Lett. \textbf{89}, 277005 (2002).

\bibitem{Scalapino:1992:PRL}
D.J.~Scalapino, S.R.~White, S.C.~Zhang, Phys. Rev. Lett. \textbf{68},
2830 (1992).

\bibitem{Peierls:1933:ZP}
R.~Peierls, Z. Phys. \textbf{80}, 763 (1933).

\bibitem{Luttinger:1951:PR}
J.M.~Luttinger, Phys. Rev. \textbf{84}, 814 (1951).

\bibitem{Nambu:1960:PR}
Y.~Nambu, Phys. Rev. \textbf{117}, 648 (1960).

\bibitem{Schrieffer:1988:Book}
J.R.~Schrieffer, \textit{Theory of Superconductivity} (Addison-Wesley,
Reading, MA, 1988).

\bibitem{Casek:2005:PRB}
P.~C\'{a}sek, C.~Bernhard, J.~Huml\'{\i}\v{c}ek, and D.~Munzar,
Phys. Rev. B \textbf{72}, 134526 (2005).

\bibitem{Vidberg:1977:JLTP}
H.J.~Vidberg and J.W.~Serene, J. Low Temp. Phys. \textbf{29}, 179 (1977).

\bibitem{Aspnes:1982:AJP}
D.~E.~Aspnes, Am.~J.~Phys. \textbf{50}, 704 (1982).

\bibitem{Ehrenreich:1966}
H.~Ehrenreich, \textit{Electromagnetic transport in solids: Optical
properties}, In \textit{The Optical Properties of Solids}, edited
by J. Tauc, Vol. 34 (Academic Press, New York, 1966).

\bibitem{Bogoljubov:1958:FdP}
N.N.~Bogoljubov, V.V.~Tolmachov, D.V.~\v{S}irkov,
Fortschritte der Physik \textbf{6}, 605 (1958).

\bibitem{Anderson:1958:PRB}
P.~W.~Anderson, Phys. Rev. \textbf{110}, 827 (1958).

\bibitem{Tinkham:1996:Book}
M.~Tinkham, \textit{Introduction to Superconductivity}
(McGraw-Hill, New York, 1996).

\bibitem{Bernhard:2008:JPCS}
C.~Bernhard, Li~Yu, A.~Dubroka, K.W.~Kim, M.~R\"{o}ssle, D.~Munzar,
J.~Chaloupka, C.T.~Lin, and Th.~Wolf,
J. Phys. Chem. Solids \textbf{69}, 3064 (2008).

\bibitem{Zelezny:2001:PRB}
V.~\v{Z}elezn\'{y}, S.~Tajima, D.~Munzar, T.~Motohashi, J.~Shimoyama,
K.~Kishio, Phys. Rev. B  \textbf{63}, 060502(R) (2001).

\bibitem{Munzar:2001:PRB}
D.~Munzar, C.~Bernhard, T.~Holden, A.~Golnik, J.~Huml\'{\i}\v{c}ek,
and M.~Cardona, Phys. Rev. B  \textbf{64}, 024523 (2001).

\bibitem{Bernhard:2000:PRB}
C.~Bernhard, T.~Holden, A.~Golnik, C.T.~Lin, M.~Cardona,
Phys. Rev. B  \textbf{62}, 9138 (2000).

\bibitem{Bernhard:YCa-123}
C.~Bernhard \textit{et al.}, unpublished.

\bibitem{Mori:2006:JofPSJ}
M.~Mori, T.~Tohyama, S.~Maekawa, J. Phys. Soc. Jpn. \textbf{75}, 034708 (2006).

\bibitem{Mori:2002:PRB}
M.~Mori, T.~Tohyama, S.~Maekawa, Phys. Rev. B \textbf{66}, 064502 (2002).

\bibitem{Lee:2007:Nature}
W.S.~Lee, I.M.~Vishik, K.~Tanaka, D.H.~Lu, T.~Sasagawa, N.~Nagaosa,
T.P.~Devereaux, Z.~Hussain, and Z.-X.~Shen, Nature \textbf{450}, 81 (2007).

\bibitem{Tanaka:2006:Science}
K.~Tanaka, W.S.~Lee, D.H.~Lu, A.~Fujimori, T.~Fujii, Risdiana, I.~Terasaki,
D.J.~Scalapino, T.P.~Devereaux, Z.~Hussain, and Z.-X.~Shen,
Science \textbf{314}, 1910 (2006).

\bibitem{Kuzmenko:2003:PRL}
A.~B.~Kuzmenko, N.~Tombros, H.~J.~A.~Molegraaf, M.~Gr{\"u}ninger, D.~van der Marel, S.~Uchida,
Phys.~Rev.~Lett \textbf{91}, 037004 (2003).

\bibitem{Basov:1999:Science}
D.N.~Basov, S.I.~Woods, A.S.~Katz, E.J.~Singley, R.C.~Dynes, M.~Xu,
D.G.~Hinks, C.C.~Homes, M.~Strongin, Science \textbf{283}, 49 (1999).

\bibitem{Norman:2000:PRB}
M.~R.~Norman, M.~Randeria, B.~Janko, J.~C.~Campuzano,
Phys.~Rev.~B \textbf{61}, 14742 (2000).

\end{thebibliography}
\end{document}